\pdfoutput=1

\documentclass[journal,10pt]{IEEEtran}

\usepackage{cite}
\usepackage{graphicx}
\usepackage{amsmath}
\usepackage{algorithmic}
\usepackage{array}
\usepackage[caption=false,font=footnotesize]{subfig}
\usepackage{url}

\usepackage{amssymb}
\usepackage{amsthm}
\usepackage{multicol}
\usepackage{multirow}
\usepackage{mdwmath}
\usepackage{mdwtab}
\usepackage{algorithm}

\begin{document}

\title{Joint QoS-Aware Scheduling and Precoding for Massive MIMO Systems via Deep Reinforcement Learning}

\author{Chih-Wei~Huang,~\IEEEmembership{Member,~IEEE}
Yen-Cheng~Chou, Hong-Yunn~Chen, and Cheng-Fu~Chou

\thanks{C.-W. Huang and Y.-C. Chou are with the Department of Communication Engineering, National Central University, Taoyuan, Taiwan (e-mail: cwhuang@ce.ncu.edu.tw, 107523031@cc.ncu.edu.tw).}
\thanks{H.-Y. Chen is with the Graduate Institute of Networking and Multimedia, National Taiwan University, Taipei, Taiwan (e-mail: d05944018@ntu.edu.tw).}
\thanks{C.-F. Chou is with the Department of Computer Science and Information Engineering, National Taiwan University, Taipei, Taiwan (e-mail: ccf@csie.ntu.edu.tw).}
\thanks{Manuscript received XXX, XX, 2021; revised XXX, XX, 20xx.}}

{}

\maketitle

\begin{abstract}
The rapid development of mobile networks proliferates the demands of high data rate, low latency, and high-reliability applications for the fifth-generation (5G) and beyond (B5G) mobile networks. Concurrently, the massive multiple-input-multiple-output (MIMO) technology is essential to realize the vision and requires coordination with resource management functions for high user experiences. Though conventional cross-layer adaptation algorithms have been developed to schedule and allocate network resources, the complexity of resulting rules is high with diverse quality of service (QoS) requirements and B5G features. In this work, we consider a joint user scheduling, antenna allocation, and precoding problem in a massive MIMO system. Instead of directly assigning resources, such as the number of antennas, the allocation process is transformed into a deep reinforcement learning (DRL) based dynamic algorithm selection problem for efficient Markov decision process (MDP) modeling and policy training. Specifically, the proposed utility function integrates QoS requirements and constraints toward a long-term system-wide objective that matches the MDP return. The componentized action structure with action embedding further incorporates the resource management process into the model. Simulations show 7.2\% and 12.5\% more satisfied users against static algorithm selection and related works under demanding scenarios.
\end{abstract}

\begin{IEEEkeywords}
Resource allocation, QoS, deep reinforcement learning, algorithm selection, massive MIMO.
\end{IEEEkeywords}

\IEEEpeerreviewmaketitle

\section{Introduction}

The rapid development of mobile networks proliferates the demands of high data rate, low latency, and high-reliability applications\cite{Saad2019}. While the traditional mobile network confronts challenges on spectrum insufficiency, the multiple-input-multiple-output (MIMO) technology, which contributes crucial progress in system capacity and reliability, is regarded as a necessary feature in the fifth-generation (5G) and beyond (B5G) wireless network systems\cite{Yang2019}. Multi-user MIMO (MU-MIMO) has been widely exploited in current wireless systems and furnishes significant improvement through conventional MIMO\cite{Alfano2020}. However, with a roughly equal number of service antennas and terminals under frequency-division duplex operation, MU-MIMO lacks scalability in various scenarios\cite{Bjornson.2017}.

The massive MIMO system\cite{Bjornson.2019,Sanguinetti.2020} has achieved breakthroughs in practice by accessing a large number of antennas on active terminals with time-division duplex\cite{Feng2016}. It is characterized by a base station (BS) equipped with more than 100 antennas that simultaneously serve multiple users with shared time-frequency resources. Extra antennas are used to steering energy in small regions to improve system throughput and energy efficiency. More advanced resource allocation is required for BSs in massive MIMO systems due to the scale of antenna numbers and resource options\cite{Chen.2018bu5}.

In massive MIMO, zero-forcing (ZF) precoding is a primary linear scheme to attain virtually optimal capacity performance taking advantage of the asymptotic orthogonality of user channels under different reflecting and scattering paths in a rich-scattering environment\cite{Liu2020}. It is generally realized through baseband processing, requiring radio frequency (RF) chains performing RF-baseband frequency transceiving and analog-to-digital conversion. The immense hardware demand limits the desired scalability coming with massive antenna array sizes. The hybrid precoding is widely adopted in recent research to mitigate hardware constraints while realizing the potential of massive MIMO systems\cite{Chen2019,Bereyhi2019}. Hence the hybrid precoding is utilized in this work.

User selection and precoding strategies in massive MIMO across media access control (MAC) and physical (PHY) layers have been actively investigated. In\cite{sheikh2019}, the authors proposed user and antenna selection algorithms to maximize the system sum-rate of a massive MIMO system with various precoding schemes. Lagen et al.\cite{Lagen2017} presented a procedure for joint user scheduling, precoder design, and transmission directing in TDD MIMO small cell networks. In\cite{Kuo2018}, an utility-based antenna allocation algorithm is proposed to efficiently allocate number of antennas to UEs in a massive MIMO system. The work considered scalable video streaming. Choi et al.\cite{Choi2018} proposed a joint user selection, power allocation, and precoder design algorithm for massive MIMO downlink systems providing gains in spectral efficiency. However, the rule-based joint precoding methods focused less on application quality of service (QoS) and require higher complexity in 5G scenarios.
With advances in artificial intelligence, deep reinforcement learning (DRL) approaches are adopted to deal with wireless network scheduling problems. Wei et al.\cite{Wei2018} proposed an actor-critic-based model for user scheduling and resource allocation to utilize radio resources in HetNets efficiently. Fiandrino et al.\cite{Fiandrino2020} also laid out a framework for machine learning (ML) based optimization for future networks. With potential to perform joint resource allocation in next-generation mobile networks, the DRL-based approaches suffer from practical learning issues\cite{Dulac-Arnold2019}. Coordinating the interaction of algorithms across scheduling and precoding functions is still an open problem.

The reinforcement learning (RL) based algorithm selection\cite{Lagoudakis2000} can be a robust framework to handle diverse QoS requirements across layers while taking advantage of established algorithms for high-performance joint adaptation. To be specific, in B5G scenarios, when a large portion of UEs is under restricts latency constraints, the systems can primarily benefit from providing higher priority to UEs with data expiring. At the instant when most UEs are traffic demanding, proportional fairness can be the preferred criteria. Though cross-layer adaptation algorithms can be developed to schedule and allocate network resources, the complexity of resulting rules increases rapidly under 5G features and diverse QoS requirements. Therefore, joint approaches adapting among feasible fundamental algorithms are worth investigating. Following the ideas of hybrid algorithm design\cite{Kotthoff2016,Kerschke2018}, the algorithm selection problem can be modeled as a Markov decision process (MDP) and solved by RL. Studies have shown that deep learning-based algorithm selection models that timely interact with environments have advantages on nonlinear and high complexity dynamic tasks\cite{Loreggia2016}. The concept was applied to 5G new radio (NR) resource allocation tasks to improve the training process but focused solely on user scheduling\cite{Tseng2019,Chen2020}.

In this work, we investigate a joint user scheduling, antenna allocation, and precoding problem in a massive MIMO system running 5G applications. The problem is evolved from conventional precoding and scheduling problems to handle strict QoS requirements from users. Instead of directly assigning resources, such as the number of antennas, the process is transformed into selecting algorithm combinations for scheduling, allocation, and precoding. An MDP model is specifically designed to resolve the dynamic algorithm selection task.
The main contributions can be summarized as follows:
\begin{itemize}
    \item We formulate a novel QoS-aware radio resource allocation problem for joint scheduling, antenna allocation, and massive MIMO precoding. The utility function integrates user requirements and constraints toward a long-term system-wide objective that matches the MDP return.
    \item A componentized MDP action structure is proposed with resource allocation functions and fundamental algorithms identified. The dynamic algorithm selection policy can thus be effectively trained.
    \item A deep deterministic policy gradient (DDPG)\cite{Lillicrap2016} based training process incorporating action embedding\cite{dulac2015} is designed to convert the action into a continuous space and take full advantage of DDPG.
\end{itemize}
The simulations are performed under realistic traffic scenarios referring to traffic types in the 5QI table~\cite{TS501}. Static algorithm combinations and baselines in the literature are implemented for comparison. Simulation results suggest that the proposed dynamic algorithm selection satisfied 7.2\% and 12.5\% more users against static algorithm selection and related joint allocation schemes under demanding scenarios.

In the rest of the paper, we first present related works on resource management and machine learning in Section~\ref{sec:related}. Section~\ref{sec:system} introduces the massive MIMO system model and problem formulation of joint scheduling and precoding. Section~\ref{sec:algorithm} describes the proposed MDP model with componentized actions. The simulation results are demonstrated in Section~\ref{sec:results}. Finally, Section~\ref{sec:conclusion} concludes the article.

\section{Related Works}\label{sec:related}

\subsection{Joint User Scheduling and Precoding}

User scheduling has been one of the primary resource allocation topics across generations of mobile communication technologies. With massive MIMO, the precoding further controls the availability of underlying physical resources and can be jointly considered for enhanced performance.
\cite{Kwon2016} presented an adaptive algorithm for joint user scheduling, precoding design, and beamforming in dynamic MIMO small cell networks. The transmit direction is optimized every frame using conventional allocation strategies across scheduling, precoding, and power control.
Based on a two-stage precoding framework for large-scale MIMO systems with frequency division duplexing, authors in\cite{Almradi2020} proposed an improved user scheduling approach with low-rank channels and precoding design. Both throughput gain and fairness were achieved.
In\cite{Saedy2017, Xie2019}, joint scheduling and precoding for matching MIMO cellular networks were investigated and analyzed.
In\cite{Olyaee2018}, the authors proposed an antenna and user selection algorithm for downlink massive MIMO transmission with ZF precoding.
Singh et al.\cite{Singh2016} developed an optimal resource fraction algorithm (ORFA) combining the proportional fair UE ranking and water filling resource allocation for MIMO networks with a minimum mean square error (MMSE) precoder.
In\cite{Kuo2018}, a utility-based layer and antenna allocation (UBLAA) algorithm is proposed to maximize the transmission efficiency for layered video streaming. The marginal utility is evaluated to determine the number of antennas assigned to UEs in a massive MIMO system.

While the joint user scheduling and precoding can be executed to some extend with conventional methods, the challenging application QoS requirements and rising complexity of 5G and beyond systems lead to performance degradation. To comprehensively integrate cross-layer functions for future networks, machine learning-based approaches are worth investigation\cite{Fiandrino2020}.

\subsection{Resource Management with Machine Learning}

Machine learning-based techniques are actively developing for next-generation network resource management. Authors in\cite{Jiang2017, Fiandrino2020} addressed the benefits of artificial intelligence-aided wireless systems and categorized primary machine learning algorithms in the context of next-generation networks. Applicable wireless communication technologies include massive MIMO, cognitive radios, heterogeneous networks, small cells, and device-to-device networks.
\cite{Li2018} built a resource management model with DRL for network slicing and demonstrated that DRL could incorporate the relation between demand and supply, enhancing network slicing agility.
\cite{Yang2018} proposed a deep reinforcement learning-enabled coverage and capacity optimization algorithm for massive MIMO systems. DRL is used to coordinate the inter-cell interference and support user scheduling dynamically.
\cite{Ye2019} also applied a DRL model for resource allocation agents in vehicle-to-vehicle (V2V) communications. The agents determine the sub-band and power levels for transmission with local observations.
Zhang et al.\cite{Zhang2020} proposed DRL-based control for resource management in spectrum sharing. With dynamic power control, both primary and secondary users efficiently meet QoS requirements.

Overall, DRL has been applied on various resource management tasks in wireless networks. However, cross-layer coordination is more complex, less studied, and requires specifically designed ML structure to be effectively solved.

\subsection{Deep Reinforcement Learning and DDPG}\label{ssec:ddpg}

In general, RL is a machine learning technique to solve decision-making problems typically modeled as an MDP, a mathematical framework to describe the target environment\cite{Sutton2018}. In RL, an agent learns through interacting with the environment and iteratively improves its ability to achieve a pre-defined goal. An MDP problem consists of states $\mathbf{s}_t\in\mathcal{S}$, actions $\mathbf{a}_t\in\mathcal{A}$, transition probabilities $Pr(\mathbf{s}_{t+1}|\mathbf{s}_t,\mathbf{a}_t)$, and rewards $r_t=r(\mathbf{s}_t,\mathbf{a}_t)\in\mathbb{R}$, where $\mathcal{S}$ and $\mathcal{A}$ are state and action spaces. In each time step $t$, an agent recognizes $\mathbf{s}_t$ from the environment and chooses a suitable $\mathbf{a}_t$. After the action is applied, the next state $\mathbf{s}_{t+1}$ and reward $r_t$ are observed from the environment. In this model, the goal is to learn the stochastic policy $\pi(\mathbf{a}_t|\mathbf{s}_t)$, which maximizes the long-term return
\begin{equation}\label{eq:return}
  R_t=\sum_{i=t}^{T}\gamma^{i-t} r(\mathbf{s}_i,\mathbf{a}_i),
\end{equation}
where $T$ and $\gamma \in (0,1)$ are termination time and the discount factor. The action-value represents the expected return when executing action $\mathbf{a}_t$ in state $\mathbf{s}_t$ following $\pi$ as
\begin{equation}
  Q^{\pi}(\mathbf{s}_t,\mathbf{a}_t)=\mathbb{E}_{\mathbf{a}\sim\pi}[R_t|\mathbf{s}_t,\mathbf{a}_t].
\end{equation}

When modeled by MDP, the massive MIMO resource allocation is a high complexity problem with large state and action spaces to present possible situations. Therefore, a deep learning-based approach, specifically DDPG\cite{lillicrap2015}, is proposed to integrate with the resource allocation process. DDPG is a landmark scheme in the policy gradient family and more suitable for applications with complex actions comparing with the deep Q-network (DQN)\cite{Mnih2015}. The deterministic policy gradient (DPG)\cite{Silver2014} concept, experience replay, slow-learning target networks from DQN, and the actor-critic structure are all integrated into DDPG.

The DDPG algorithm utilize the recursive Bellman equation to evaluate action-value functions. Thus the deterministic policy $\mu:\mathcal{S}\rightarrow \mathcal{A}$ provides the action $\mathbf{a}_t=\mu(\mathbf{s}_t)$, and the action-value function becomes
\begin{equation}\label{eq:action-value}
  Q(\mathbf{s}_t,\mathbf{a}_t)=\mathbb{E}_{r_t,\mathbf{s}_{t+1}\sim E}[r(\mathbf{s}_t,\mathbf{a}_t)+\gamma Q(\mathbf{s}_{t+1},\mu(\mathbf{s}_{t+1}))].
\end{equation}
Furthermore, considering function approximators parameterized by $\theta^Q$ and $\theta^\mu$, we can optimize the action-value function by minimizing the loss function. The actor network updates the policy with aids from the critic network, where the policy gradient is\cite{Silver2014}
\begin{equation}\label{eq:policy_gradient}
    \begin{aligned}
        \nabla_{\theta^\mu}J&\approx\mathbb{E}_{r_t,\mathbf{s}_{t+1}\sim E}[\nabla_{\theta^\mu}Q(\mathbf{s}_t,\mathbf{a}_t)]\\
        &=\mathbb{E}_{\mathbf{s}_t\sim E}[\nabla_{\mathbf{a}}Q(\mathbf{s}_t,\mathbf{a}_t|\theta^Q)\nabla_{\theta_\mu}\mu(\mathbf{s}_t|\theta^\mu)].
    \end{aligned}
\end{equation}
Accordingly, the training procedures using samples from experience replay, $E$, can be realized.

\section{System Model and Problem Formulation}\label{sec:system}

This section describes the massive MIMO network structure and problem formulation. A joint user scheduling, antenna allocation, and precoding problem is proposed with the potential to be modeled as an MDP.

\subsection{System Model}

We consider a single-cell massive MIMO system consisting of an $M$-antenna BS and $K$ single-antenna user equipments (UEs). Thus we have $m\in\mathbf{M}=\{1,2,\dots,M\}$ and $k\in\mathbf{K}=\{1,2,\dots,K\}$. During transmission, the BS allocates $N_{k,t}$ number of antennas to UE $k$ at the $t$-th transmission time interval (TTI), where a TTI is $T_I$-second long. Each UE is associates with a type of traffic, referring to the 5QI table\cite{3gpp23501}. Considered UE properties include channel quality indicators, requested data, and a traffic type. The channel quality indicator, $\mathbf{CQI}$, can be obtained from the table defined in\cite{3gpp38214} given the signal to interference and noise ratio (SINR). Requested data, $\mathbf{D}_k$, is the set of data packets generated for transmission. The properties attached with a traffic type, $\mathbf{TYPE}$, are packet size, mean packet arrival time, latency constraint, guarantee bit-rate, and error rate constraint.

\begin{figure}
    \centering
    \includegraphics[width=8.5cm]{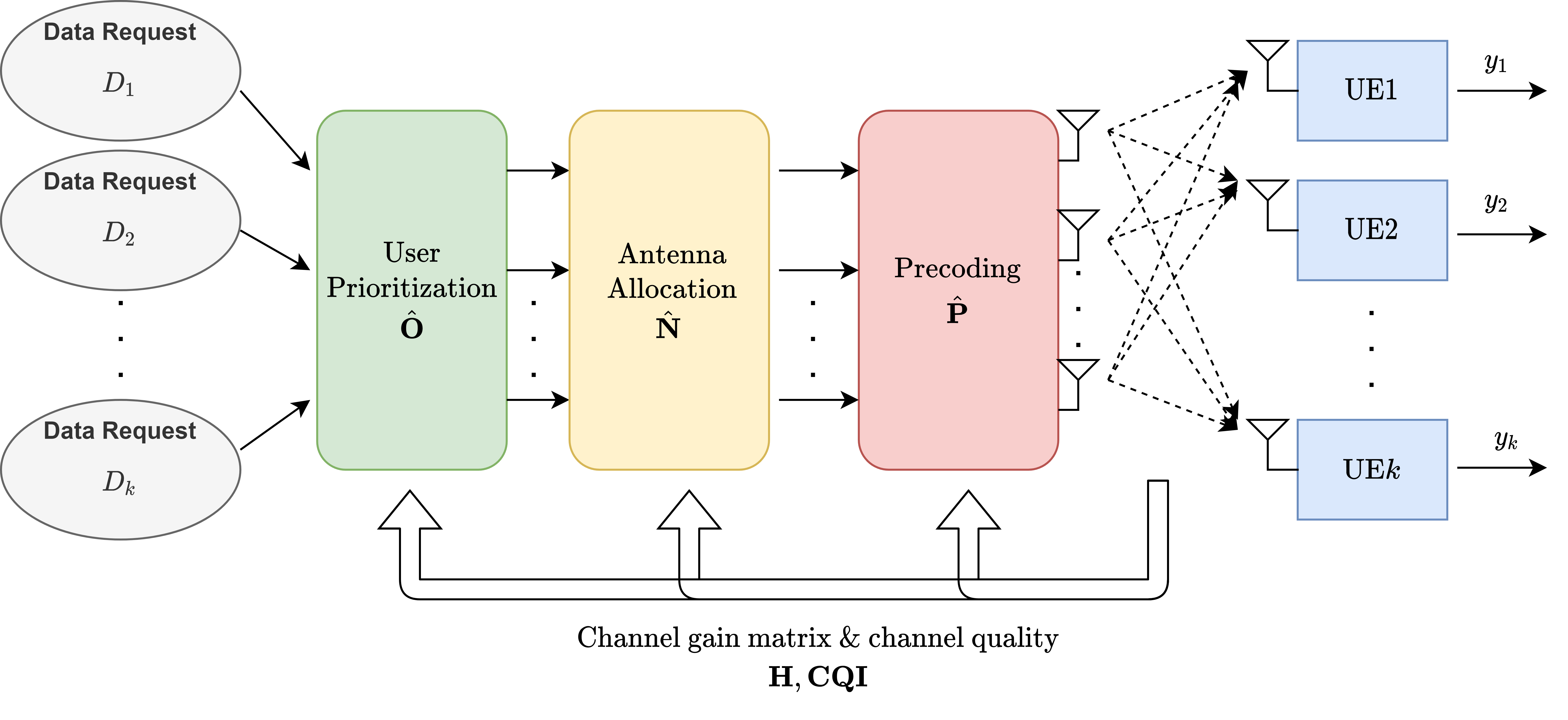}
    \caption{Joint scheduling and precoding massive MIMO model in three function components: user prioritization, antenna allocation, and precoding.}
    \label{fig:model}
\end{figure}

The joint user resource allocation and precoding model is illustrated in Figure~\ref{fig:model}. The procedure consists of three \emph{function components}, including user prioritization, antenna allocation, and precoding. The outcome of user prioritization is defined as $\hat{\mathbf{O}} = \{\mathbf{O}_t|1\leq t\leq T\}$, which ranks UEs every TTI. $\mathbf{O}_t$ is an \emph{ordered} subset of UEs containing ones that have requested data to be transmitted at $t$. The antenna allocation results, $\hat{\mathbf{N}}=\{\mathbf{N}_t|1\leq t\leq T\}$, record the number of antennas assigned to prioritized UEs, where $\mathbf{N}_t=\{N_{k,t}|k\in \mathbf{O}_t\}$. The precoding matrix set, $\hat{\mathbf{P}}=\{\mathrm{P}_t|1\leq t\leq T\}$, is the set of precoding matrix given antenna allocation, i.e., $\mathrm{P}_t\equiv \mathrm{P}(\mathbf{N}_t)$.

Given user data requests, $\mathbf{D}_k$, the policies of function components are determined according to channel quality feedbacks $\mathrm{H}$ and $\mathbf{CQI}$. Finally, a UE decodes the received signal $y_k$ for the data.

\begin{table}
\caption{Summary of Notations}
\centering
\small
\begin{tabular}{|p{1.5cm}p{6.0cm}|}
\hline
\textbf{Notation} & \textbf{Description} \\ \hline
$t$ & Index of the $t$-th TTI \\
$\mathbf{M}/M$ & BS antenna set / total number of antennas\\
$\mathbf{K}/K/K_r$ & UE set / total number of UEs / simultaneously receiving UEs \\
$\mathbf{\hat{O}} / \mathbf{O}_t$ & Ordered UE set / the ordered UE set at $t$ \\
$\mathbf{\hat{N}} / \mathbf{N}_t / N_{k,t}$ & Antenna allocation set / the antenna allocation set at $t$ / number of antennas allocated to UE $k$ at $t$\\
$\mathbf{\hat{P}} / \mathrm{P}_t / \mathbf{p}_{k,t}$ & Precoding matrix set / the precoding matrix at $t$ / $k$-th column vector of $\mathrm{P}_t$ \\
$\mathrm{H} / \mathbf{h}_{k,t}$ & Channel matrix / channel vector of UE $k$ at $t$ \\
$\Phi_{k,t}$ & Data transmission rate of UE $k$ at $t$\\
$U_k / U_{k,t}$ & Overall utility of UE $k$ / utility of UE $k$ up to $t$ \\
$\nu_{k,t}$ & Number of successfully received packets by UE $k$ at $t$ \\
$E_k$ & Packet error rate requirement of UE $k$ \\
$\mathbf{D}_k$ & Set of requested data packets of UE $k$ \\
$\mathbf{t}(d)$ & TTIs assigned to transmit data packet $d$ in time \\
$u^d_{k,t}$ & Receiving status of data packet $d$ of UE $k$ up to $t$ \\
$\varepsilon_d$ & Packet size of the data $d$ \\
$W$ & System bandwidth \\
$\rho$ & BS transmission power \\
$y_k$ & Received signal of UE $k$ \\
$\mathbf{x}$ & Transmitted signal for UE  \\
$\mathbf{n}$ & The additive white Gaussian noise \\
$T_I$ & The duration of a TTI \\
$T$ & Termination time in number of TTIs \\
$B$ & Number of transitions in a mini-batch \\
\hline
\end{tabular}
\label{T:notation}
\end{table}

\subsection{Hybrid Precoding}

The precoding influences the spectral efficiency by evaluating the precoding matrix and providing raw capacity for resource allocation\cite{Guo2020}. Assuming there are $K_r$ UEs simultaneously receiving data at a TTI, the received signal vector $\mathbf{y}=[y_1,y_2,\dots,y_{K_r}]^T \in \mathbb{C}^{K_r \times 1}$ can be expressed as
\begin{equation}
  \mathbf{y} = \mathrm{H}\mathbf{x} + \mathbf{n}.
\end{equation}
$\mathrm{H}\in\mathbb{C}^{K_r\times M}$ denotes the channel matrix of all $K_r$ users; $\textbf{h}_k\in\mathbb{C}^{1 \times M}$ is user $k$'s channel vector. $\mathbf{n} \sim \mathcal{CN}(0,\sigma^2)$ is the additive white Gaussian noise (AWGN) with variance $\sigma^2$, $\mathbf{n}=[n_1,n_2,\dots,n_{K_r}]^T$. Processed by a power amplifier, the transmitted signal vector $\textbf{x}\in \mathbb{C}^{M \times 1}$ is transmitted through the antennas and is given as
\begin{equation}
  \mathbf{x}=\mathrm{P}\chi.
\end{equation}
$\mathrm{P}=[\mathbf{p}_1,\mathbf{p}_2,\dots,\mathbf{p}_{K_r}]\in\mathbb{C}^{M\times K_r}$ with column vectors $\textbf{p}_k\in\mathbb{C}^{M \times 1}$, is the set of hybrid precoding matrices. $\mathbf{\chi}=[\chi_1,\chi_2,\dots,\chi_{K_r}]^T \in \mathbb{C}^{K_r \times 1}$ is the modulated user signals with $\mathbb{E}[\chi\chi^H] =(\frac{\rho}{K_r})\mathrm{I}_{K_r}$, where $\mathrm{I}$ and $\rho$ refer to the unit matrix and the total transmission power. With the system model defined in Figure~\ref{fig:model}, the received signal of user $k$ is
\begin{equation}
    y_k=\mathbf{h}_k\sum^{K_r}_{i=1}\mathbf{p}_i\chi_i+n_k.
\end{equation}

Therefore, after utilizing resource allocation functions, we have $K_r=\|\mathbf{O}_t\|$ with the time index $t$. The signal-to-interference-plus-noise ratio (SINR) of UE $k$ at $t$ is\cite{GAO201779}
\begin{equation}
  SINR_{k,t}=\frac{\frac{\rho}{\|\mathbf{O}_t\|}|\mathbf{h}_{k,t}\mathbf{p}_{k,t}|^2}{\sigma^2+\frac{\rho}{\|\mathbf{O}_t\|}\sum_{j\in \mathbf{O}_t,j\neq k}|\mathbf{h}_{k,t}\mathbf{p}_{j,t}|^2}.
\end{equation}
$\mathrm{P}_t\in\mathbb{C}^{M\times\|\mathbf{O}_t\|}$ is a function of antenna allocation $\mathbf{N}_t$ with column vectors $\mathbf{p}_{k,t}\in\mathbb{C}^{M\times 1}$. The data transmission rate $\Phi$ of UE $k$ at time $t$ is
\begin{equation}\label{eq:transmission_rate}
  \Phi_{k,t} = W\cdot\log\left(1+SINR_{k,t}\right),
\end{equation}
where $W$ is the system bandwidth. We formulate the precoding problem as
\begin{equation}\label{eq:precoding_problem}
    \max_{\mathrm{P}_t}\sum_k \Phi_{k,t},
\end{equation}
subject to
\begin{equation}\label{eq:precoding_c1}
    \sum_{k\in\mathbf{O}_t} p^m_{k,t} \leq 1, \quad\forall m \in \mathbf{M}.
\end{equation}
$p^m_{k,t}$ is an element of $\mathbf{p}_{k,t}$, and \eqref{eq:precoding_c1} is the constraint of the precoding matrix gain.

\subsection{User Scheduling and Antenna Allocation}

In general, user resource allocation aims to maximize the total system utility by actively distributing resources. Packet-level transmission utility is first defined to describe the QoS status in terms of requested data. Assuming set $\mathbf{t}(d)$ is the TTIs assigned to transmit data packet $d$ within its latency constraint. $u^d_{k,t}$ indicates the receiving status of data packet $d\in\mathbf{D}_k$ and is defined as
\begin{equation}
    u^d_{k,t} =
    \begin{cases}
    1, & \text{if } \sum_{i\in \mathbf{t}(d)}\Phi_{k,i}\cdot T_I \geq \varepsilon_d\\
    0, & \text{otherwise}.
    \end{cases}
\end{equation}
A packet is successfully received, i.e., $u^d_{k,t}=1$, if sufficient resources are allocated to a packet in time. $\varepsilon_d$ is the packet size. Consequently, the number of successfully received packets by UE $k$ up to time $t$ is
\begin{equation}
  \nu_{k,t}=\sum_{d\in \mathbf{D}_{k}} u^d_{k,t}.
\end{equation}

Based on the receiving status mentioned above, the user resource allocation problem is defined to maximize the total utility every TTI. The utility gain of UE $k$ up to the $t$-th TTI, $U_{k,t}$, is a function of transmission data rates allocated to it over time. Simultaneously, application requirements, including guarantee bit rate (GBR), packet loss rate, and latency, are subjected to be satisfied. The allocation decision prioritizes UEs for $\mathbf{O}_t$ and determines the corresponding number of antennas $\mathbf{N}_t$. We formulate the problem as
\begin{equation}\label{eq:scheduling_problem}
  \max_{\mathbf{O}_t,\mathbf{N}_t}\sum_{k} U_{k,t}(\Phi_{k,i}|1\leq i\leq t),
\end{equation}
subject to
\begin{subequations}
\begin{equation}\label{eq:sp_c1}
  \frac{1}{t}\sum^t_{i=1}\Phi_{k,i} \geq GBR_k, \quad\forall k \in \mathbf{K}
\end{equation}
\begin{equation}\label{eq:sp_c2}
  1-\frac{\nu_{k,t}}{\|\mathbf{D}_{k}\|} \leq E_k, \quad\forall k \in \mathbf{K}
\end{equation}
\begin{equation}\label{eq:sp_c3}
  \sum_{k\in \mathbf{O}_t} N_{k,t} \leq M.
\end{equation}
\end{subequations}
$E_k$ is the packet error rate requirement from UE $k$'s traffic type. \eqref{eq:sp_c1} shows the derivation and constraint of providing GBR for $k$. \eqref{eq:sp_c2} is the packet error rate constraint. \eqref{eq:sp_c3} limits the total number of antennas allocated. The utility function is open to be further defined.

\subsection{QoS-Aware Joint Resource Allocation}\label{ssec:joint_problem}

It is challenging to adapt options from precoding to scheduling effectively. Problems~\eqref{eq:precoding_problem} and~\eqref{eq:scheduling_problem} are inter-dependent with different objectives. Given the antenna allocation, the precoding matrix determines the resulting throughput, while the antenna resources are allocated through user scheduling based on throughput-dependent utility. We model the complex interaction with a utility function integrating requirements and dependencies toward a long-term system objective. Under the componentized structure and adaptive algorithm selection, the problems are jointly processed.

The QoS-aware joint resource allocation objective is to maximize the number of satisfied users in the system given their application requirements. Therefore, we propose to redefine the general utility objective with requirements integrated up to the termination time $T$. The received utility of UE $k$, $U_k$, is set to 1 when GBR, loss, and latency requirements are all satisfied given allocated antenna resources. The utility function can be expressed as
\begin{equation}\label{eq:utility}
  U_k \equiv U_{k,T} =
  \begin{cases}
    1, & \text{if } 1-\frac{\nu_{k,T}}{\|\mathbf{D}_k\|} \leq E_k \\
    & \quad \land \frac{1}{T}\sum^T_{i=1}\Phi_{k,i} \geq GBR_k\\
    0, & \text{otherwise}.
  \end{cases}
\end{equation}
The joint problem is formulated by QoS requirements embedded utility and resource constraints as
\begin{equation}\label{eq:problem}
    \max_{\hat{\mathbf{O}},\hat{\mathbf{N}},\hat{\mathbf{P}}}\sum_{k} U_{k},
\end{equation}
subject to
\begin{subequations}
\begin{equation}\label{eq:c1}
  \sum_{k\in \mathbf{O}_t} N_{k,t} \leq M, \quad 0\leq t\leq T
\end{equation}
\begin{equation}\label{eq:c2}
  \sum_{k\in\mathbf{O}_t} p^m_{k,t} \leq 1, \quad\forall m \in \mathbf{M}.
\end{equation}
\end{subequations}
The objective is to maximize the total number of satisfied UEs by determining the optimal $\hat{\mathbf{O}}, \hat{\mathbf{N}}, \hat{\mathbf{P}}$ over time. \eqref{eq:c1} limits the total numbers of the allocated antennas. \eqref{eq:c2} is the constraint of the precoding matrix gain. The problem features a utility function depending on complex criteria and long-term returns. Therefore, an MDP-based solution, which models complex agent-environment interaction and optimizes future return during the process, adequately fits the problem.

\section{Deep Reinforcement Learning for Joint Massive MIMO Resource Allocation}\label{sec:algorithm}

In this section, we formulate the MDP problem with states, actions, and rewards. Also, the resource allocation function components and the DDPG training procedures are detailed.

\subsection{Markov Decision Process Formulation}

\begin{figure}
    \centering
    \includegraphics[width=8.5cm]{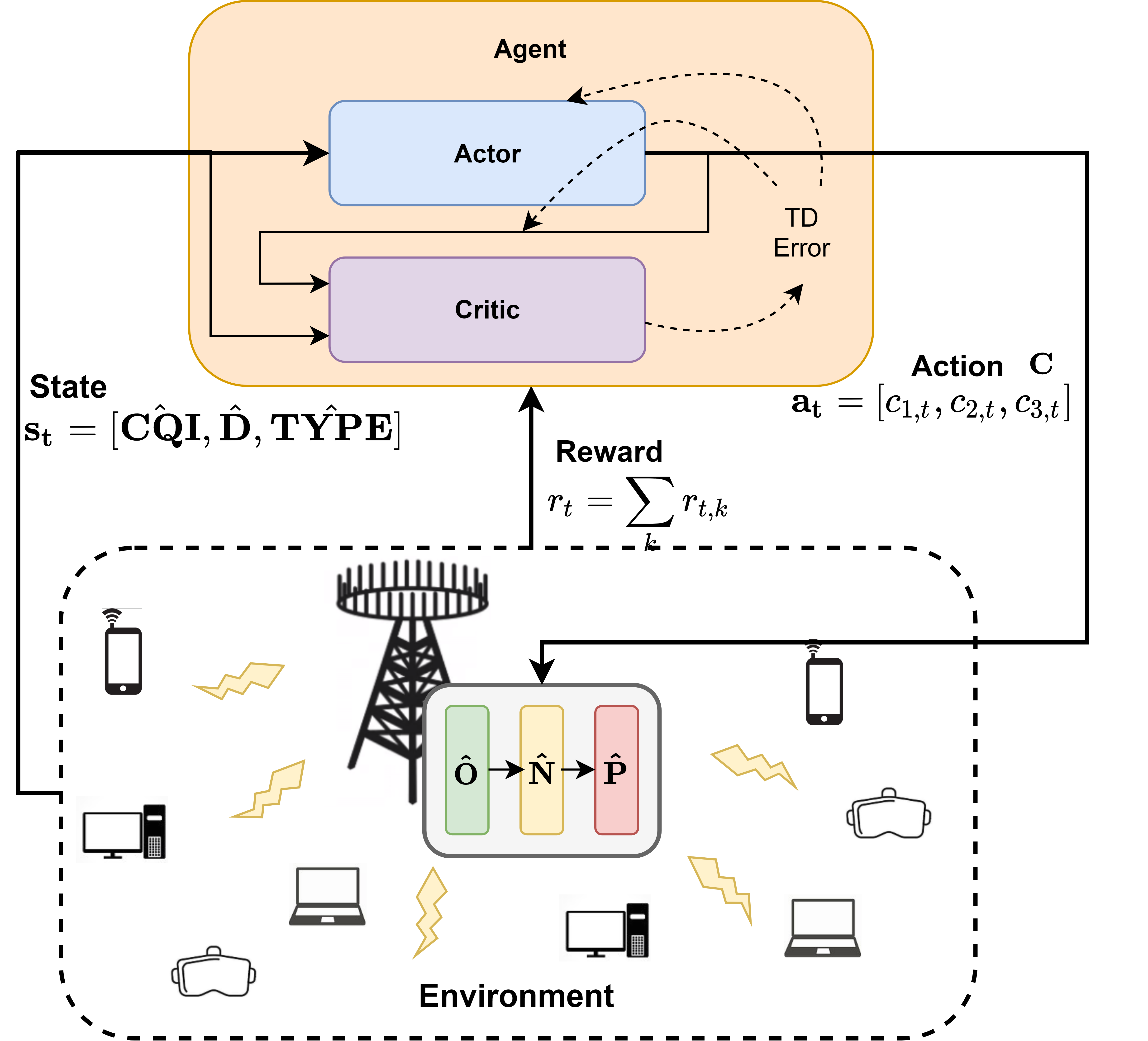}
    \caption{DDPG structure for massive MIMO resource allocation.}
    \label{fig:DDPG}
\end{figure}

\begin{figure}
    \centering
    \includegraphics[width=8.5cm]{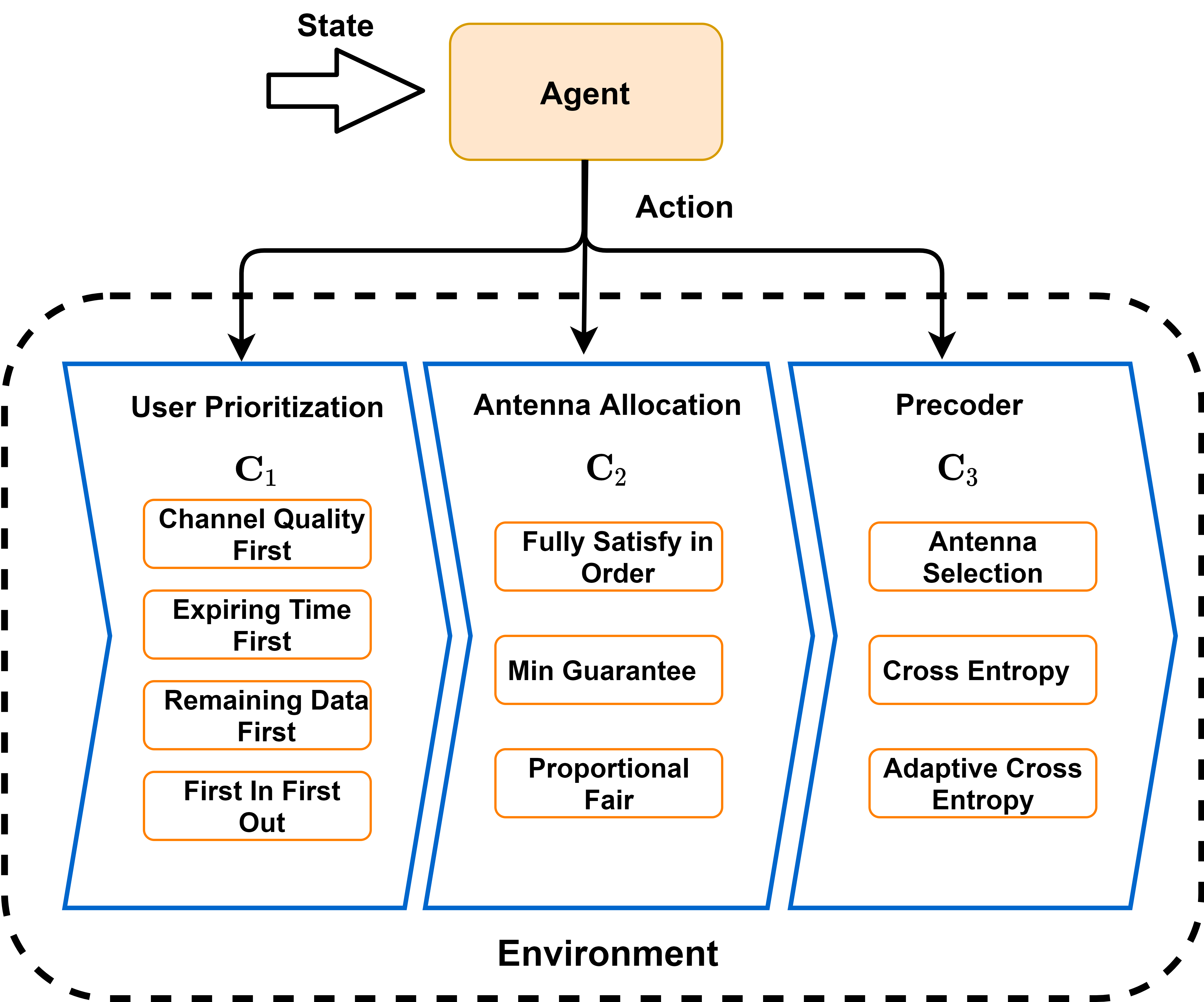}
    \caption{Componentized actions and fundamental methods in each component.}
    \label{fig:actions}
\end{figure}

Figure~\ref{fig:DDPG} illustrates the massive MIMO resource allocation problem in the DDPG structure.
During the RL process, the control agent collects \emph{state} information to determine resource allocation actions. The information includes the sets of UE channel quality $\hat{\mathbf{CQI}}$, UE data requests $\hat{\mathbf{D}}$, and traffic types $\hat{\mathbf{TYPE}}$. Thus, the state at the $t$-th TTI is defined as
\begin{equation}
    \mathbf{s}_t = [\hat{\mathbf{CQI}}, \hat{\mathbf{D}}, \hat{\mathbf{TYPE}}].
\end{equation}
Based on the problem formulated in Section~\ref{ssec:joint_problem}, the resource allocation \emph{action} is formed as a combination of components: user prioritization, antenna allocation, and precoder selection. Fundamental schemes proven to be helpful in specific scenarios are included in a component. The componentized architecture is shown in \figurename~\ref{fig:actions}. The action dynamically selects a scheme in each component according to the state observed and is expressed as
\begin{equation}
    \mathbf{a}_t=[c_{1,t}, c_{2,t}, c_{3,t}].
\end{equation}
The details of included components are described later in Section~\ref{ssec:actions}.

The reward is designed to keep the data transmission on pace considering traffic type-specific GRB and latency requirements. Due to higher uncertainty to quantify the advantage of proactive transmission\cite{Mao2016}, we adopt negative rewards to discourage situations with transmission progress falling behind. The reward function is formulated as
\begin{equation}\label{eq:reward_kt}
    r_{k,t}=\left(\frac{\nu_{k,t}}{\|\mathbf{D}_k\|}-1\right)\left(1+\alpha\cdot\left(1-\frac{\sum^t_{i=1} \Phi_{k,i}}{t\cdot GBR_k}\right)\right).
\end{equation}
The first term reflects the incompletion ratio of requested data up to time $t$. As a penalty, the value is negative if requested data from UE $k$ are not fully transmitted. If all request data are transmitted, the first term and thus the reward $r_k$ becomes zero. The second term is the adjustment to keep the transmission data rate $\Phi$ on the pace of $GBR_k$. $\alpha$ is the penalty weight, and $\alpha =0$ when the traffic type has no GBR assigned. Therefore, the reward function is
\begin{equation}
    r_t=\sum_{k}r_{k,t},
\end{equation}
which is recorded in the training process to maximize the future return, $R_t$. With the reward function design and MDP-based future return optimization, the learning process is fully aligned with the utility optimization problem~\eqref{eq:problem}.

\subsection{Componentized Actions}\label{ssec:actions}

The componentized action is the concept introduced to facilitate dynamic resource allocation via algorithm selection and improve DDPG training. As shown in \figurename~\ref{fig:actions}, we decompose the scheduling and precoding process into three function components; a component contains several \emph{fundamental methods} as algorithm options. The included algorithms are diverse in design concepts for meaningful selection. Component 1 ($\mathbf{C}_1$) prioritize UE according to specific criteria. Component 2 ($\mathbf{C}_2$) decides the number of antennas allocated to each UE per TTI. The hybrid precoding method is determined in Component 3 ($\mathbf{C}_3$). The adopted fundamental methods are introduced as follows.

The UE prioritization component ranks UEs in the system. Four implemented sorting methods are:
\begin{itemize}
    \item Channel quality first (CQI): sorts UEs according to channel conditions. A UE with higher channel quality is ranked higher.
    \item Expiring time first (Delay): ranks UEs on how close its oldest requested data is expired. It depends on the traffic-specific time constraints and how long the transmission is being delayed.
    \item Remaining data first (Remain): sorts UEs according to the size of requested data remaining in the queue, i.e., $\|\mathbf{D}_{k,t}\|-\nu_{k,t-1}$. A UE receives higher priority with more untransmitted data.
    \item First-in-first-out (FIFO): prioritizes UEs with the arrival time of the earliest arrival packet.
\end{itemize}
Thus, component $c_{1,t}\in \mathbf{C}_1=\{\text{CQI}, \text{Delay},\text{Remain},\text{FIFO}\}$. An ordered UE set $\mathbf{O}_t$ is generated every TTI.

The second component is to allocate the system resources, i.e., the number of antennas $N_{k,t}$, to UEs based on the ordered set $\mathbf{O}_t$. As a result, $c_2$ also controls the final number of UE, which can be granted a transmission opportunity. In addition to fundamental methods, a percentage parameter $\iota$ is also integrated to extend the options. The fundamental allocation methods implemented are:
\begin{itemize}
    \item Fully satisfy in order (FSO): allocates sufficient numbers of antennas to fully transmit the remaining requested data of each UE, $\|\mathbf{D}_{k}\|-\nu_{k,t-1}$, in the order of $\mathbf{O}_t$ until exhausting the system resource. The number of antennas to fully satisfy a UE, $N^{fs}_{k,t}$, is defined as
    \begin{equation}
        N^{fs}_{k,t}\equiv(N_{k,t}|\Phi_{k,t}\cdot T_I\geq(\|\mathbf{D}_{k}\|-\nu_{k,t-1})\cdot\varepsilon_k).
    \end{equation}
    \item Minimum guarantee (MinG)\cite{Guan2011}: evenly distributes a portion of antennas to a subset of UEs, $\mathbf{O}^G_t\subseteq\mathbf{O}_t$, and applies FSO on the remaining resources. Therefore, several UEs can receive a minimum share of antennas and the portion of resources reserved for even distribution, $\iota^G$, is a key parameter to consider. We determine the number of UEs receiving guaranteed resources according to the smallest $N^{fs}_{k,t}$ and can be expressed as
    \begin{equation}
        \|\mathbf{O}^G_t\|=\frac{\iota^G\cdot M}{\min_{k\in\mathbf{O}_t} N^{fs}_{k,t}},
    \end{equation}
    where $\iota^G=\{25\%,50\%,75\%,100\%\}$. Thus, there are four MinG-based options in $\mathbf{C}_2$. For example, the option with $\iota^G=50\%$ is denoted as $\mathrm{MinG50}$.
    \item Proportional fair (PF)\cite{Kim2005}: considers a subset of UEs and allocates antenna resources proportional to the ratio of currently available data rate, $\Phi_{k,t}$, to historical transmission rate. In practice, the historical transmission rate can be updated through moving averages. The parameter $\iota^{pf}=\{ 25\%, 50\%, 75\%, 100\%\}$ determines the percentage of UEs in $\mathbf{O}_t$ to be included.
\end{itemize}
With all the fundamental schemes and parameters, the complete option set $\mathbf{C}_2$ has nine elements.

For high spectrum efficiency in massive MIMO transmission, the third component selects a hybrid precoding algorithm to evaluate the precoding matrix. The fundamental hybrid precoders are:
\begin{itemize}
    \item Antenna selection (AS)\cite{Mendez2015}: greedily chooses antennas to achieve high single antenna efficiency.
    \item Cross entropy (CE)\cite{rubinstein2013}: is a probabilistic model-based algorithm iteratively solving the combining problem. The algorithm computes the achievable sum-rate of each candidate and selects the best candidates as "elites." Base on the selected elites, the probability distribution is updated by minimizing the cross entropy. CE precoding performs well with sufficient resources and a less saturated system.
    \item Adaptive cross entropy (ACE)\cite{Gao2017}: is a variation based on the CE algorithm. The ACE algorithm weights "elites" adaptively based on its achievable sum-rates. This precoding method can gain better SINR than CE in saturated situations.
\end{itemize}
The component $c_{3,t}\in \mathbf{C}_3=\{\text{AS}, \text{CE}, \text{ACE}\}$.
Overall, the action $\mathbf{a}_t$ can be one of 108 component combinations with all options considered.

\subsection{Action Embedding and Training Procedures}

As introduced in Section~\ref{ssec:ddpg}, DDPG takes advantage of DQN, DPG, and the actor-critic structure\cite{Lillicrap2016}; it is utilized to make resource allocation decisions for our target MDP problem with continuous or high dimensional states and actions. We extend the actions in this work to a continuous space through action embedding\cite{dulac2015}, where the original discrete actions are embedded in continuous upon which the actor can generalize.
The function $\upsilon:\mathbb{R}^{\mathrm{dim}(\mathcal{A})}\rightarrow \mathcal{A}$ is defined to convert the \emph{continuous} action $\mathbf{\check{a}}_t$ used for training into the \emph{discrete} action $\mathbf{a}_t$ applied to the environment, with $\mathrm{dim}(\mathcal{A})$ denoting the dimension of action space $\mathcal{A}$. Therefore, the converting function is expressed as
\begin{equation}\label{eq:converting_function}
  \mathbf{a}_t=\upsilon(\mathbf{\check{a}}_t),
\end{equation}
where $\mathbf{\check{a}}_t=[\check{c}_{1,t},\check{c}_{2,t},\check{c}_{3,t}]$ is the action formed by continuous component values. Also, the deterministic policy generating continuous action $\check{\mu}:\mathcal{S}\rightarrow \mathbb{R}^{\mathrm{dim}(\mathcal{A})}$ is applied in the model as the actor network $\check{\mu}(\mathbf{s}_t|\theta^\mu)$.

The training process is described in Algorithm~\ref{alg:DDPG_procedure}.
First, networks are initialized.
Every TTI, the agent generates continuous action $\mathbf{\check{a}}_t=\check{\mu}(\mathbf{s}_t|\theta^\mu)+\mathcal{N}_t$ from the actor with random noise $\mathcal{N}_t$ for exploration.
The discrete action $\mathbf{a}_t$ is obtained from \eqref{eq:converting_function} and applied to the environment for the reward $r_t$ and the next state $\mathbf{s}_{t+1}$ as feedbacks. In order to reuse execution experiences, DDPG stores transition $(\mathbf{s}_t, \mathbf{a}_t, \mathbf{s}_{t+1}, r_t)$ in the replay buffer. After that, DDPG samples $B$ number of transitions from the replay buffer to form a mini-batch $\mathcal{B}$. With mini-batch inputs, the target actor network $\check{\mu}'(\mathbf{s}_{t+1}|\theta^{\mu'})$ outputs the action to the target critic network $Q'$, where the resulting action-value can be evaluated based on~\eqref{eq:action-value}. Therefore, the critic network is updated by minimizing the loss function
\begin{multline}\label{eq:embedded_loss}
  L(\theta^Q)=\frac{1}{B}\sum_{i\in\mathcal{B}}\left[r_i+\gamma Q'(\mathbf{s}_{i+1},\upsilon(\check{\mu}'(\mathbf{s}_{i+1}|\theta^{\mu'}))|\theta^{Q'})\right. \\
  \left.\vphantom{\sum}-Q(\mathbf{s}_i,\mathbf{a}_i|\theta^Q)\right].
\end{multline}
The actor network is updated following the deterministic policy gradient theorem modified from \eqref{eq:policy_gradient} as\cite{dulac2015}
\begin{multline}\label{eq:actor_update}
  \nabla_{\theta^\mu}J(\theta^\mu)\approx \\
  \frac{1}{B}\sum_{i\in\mathcal{B}}\nabla_{\mathbf{\check{a}}} Q(\mathbf{s}_i,\check{\mu}(\mathbf{s}_i|\theta^\mu)|\theta^Q)\nabla_{\theta^\mu}\check{\mu}(\mathbf{s}_i|\theta^\mu).
\end{multline}
We note that the replay buffer stores the discrete action generated by \eqref{eq:converting_function}, but the policy gradient is taken at $\check{\mu}$. This allows the learning algorithm to leverage action executed to the environment for critic network training, while taking the policy gradient at the actual output of the actor network. Finally, DDPG uses the soft-update to improve both critic and actor target networks with the constant $\tau$ as
\begin{equation}\label{eq:soft_update}
  \begin{aligned}
    \theta^{Q'} &\leftarrow \tau\theta^Q + (1-\tau)\theta^{Q'},\\
    \theta^{\mu'} &\leftarrow \tau\theta^\mu + (1-\tau)\theta^{\mu'}.
  \end{aligned}
\end{equation}
The parameters in target networks change slowly and considerably improve the learning stability.

\begin{algorithm}
  \caption{The DDPG Training with Action Embedding}
  \label{alg:DDPG_procedure}
  \begin{algorithmic}[1]
    \STATE Randomly initialize critic network $Q$ and actor network $\check{\mu}$ in the DDPG agent
    \STATE Initialize target network $Q'$ and $\check{\mu}'$ with weights $\theta^{Q'}\leftarrow\theta^Q,\theta^{\mu'}\leftarrow\theta^\mu$.
    \STATE Initialize replay buffer
    \FOR{episode$=1$ to end}
      \STATE Initialize a random process $\mathcal{N}$ for action exploration
      \STATE Receive initial observation state $\mathbf{s}_1$
      \FOR{t=1, T}
        \STATE Generate continuous action $\mathbf{\check{a}}_t=\check{\mu}(\mathbf{s}_t|\theta^\mu)+\mathcal{N}_t$ from actor in DDPG
        \STATE Convert the action form continuous to discrete $\mathbf{a}_t=\upsilon(\mathbf{\check{a}}_t)$ to embedding on three components $[c_{1,t}, c_{2,t}, c_{3,t}]$
        \STATE Execute action $\mathbf{a}_t$ and observe reward $r_t$ and new state $\mathbf{s}_{t+1}$
        \STATE Store transition $(\mathbf{s}_t,\mathbf{\check{a}}_t,r_t,\mathbf{s}_{t+1})$ in replay buffer
        \STATE Sample a random mini-batch from the replay buffer
        \STATE Update the critic by minimizing the loss \eqref{eq:embedded_loss}
        \STATE Update the actor using the gradient \eqref{eq:actor_update}
        \STATE Update the actor and critic network with the equation~(\ref{eq:actor_update})(\ref{eq:soft_update})
      \ENDFOR
    \ENDFOR
  \end{algorithmic}
\end{algorithm}

\section{Numerical Results}\label{sec:results}

This section introduces simulation settings for traffic scenarios, the massive MIMO environment, and DDPG training. Numerical results compare the proposed learning-based method with baselines, including static combinations of fundamental methods and related works.

\subsection{Simulation Setup}

The simulation scenarios are built as mixes of applications in a massive MIMO system. Table~\ref{table:traffic_type} shows six selected traffic types based on 5QI specifications\cite{3gpp23501}, including voice over IP (VoIP), video streaming, gaming, and virtual reality (VR) / augmented reality (AR). The properties attached to a traffic type include latency requirement, GBR, packet size, mean packet arrival time, and error rate requirement. A UE is admitted to the system as a traffic session with a predetermined type and properties to generate requested data. For scenarios, traffic sessions from all types are mixed in various UE ratios listed in Table~\ref{table:traffic_scenarios} with specific focuses. For the communication system, COST2100\cite{Liu2012} is used to model the MIMO channel, and UEs are distributed following the Poisson point process (PPP).

The simulation datasets are form by 60000-TTI-long data blocks containing CQIs and requested data of UEs every TTI. We generate four data blocks for each of six traffic types for training, resulting in 24 distinguish traffic data blocks. In an epoch, the training goes through 24 data blocks in random orders. The model converges after 72 to 74 epochs of training. Therefore the resulting model is expected to handle traffic scenarios in an arbitrary mix of data types. The testing is performed on ten separately generated data blocks for each scenario. Also, the penalty weight $\alpha$ in \eqref{eq:reward_kt} is set to 0.5 when GBR is available. The continuous component values in \eqref{eq:converting_function} are set in $[-1,1]$ and evenly distributed for discrete actions with $\mathrm{dim}(\mathcal{A})=3$.
Table~\ref{table:system_parameters} shows the complete parameter list. The training and decision-making models are implemented using TensorFlow\cite{tensorflow2015-whitepaper} version 1.14 on a desktop machine with Intel i7-3770 CPU and Nvidia RTX 2080Ti GPU under parameters listed in Table~\ref{table:ML_parameters}.

\begin{table}
\renewcommand{\arraystretch}{1.3}
\caption{Traffic Type Parameters}
\label{table:traffic_type}
\centering
\scriptsize
\begin{tabular}{|p{0.5cm}|p{1.2cm}|p{0.7cm}|p{0.8cm}|p{0.7cm}|p{0.6cm}|p{0.7cm}|}
\hline
\textbf{Type (5QI Value)} & \textbf{Application} & \textbf{Packet Size (Bytes)} & \textbf{Mean Packet Arrival Time (ms)} & \textbf{Latency (ms)} &\textbf{GBR (Mb/s)} & \textbf{Error Rate}\\\hline
A (1) & VoIP & 200 & 15 & 100 & 0.112 & $10^{-2}$\\ \hline
B (2) & Video & 1250 & 5 & 150 & 0.8 & $10^{-3}$\\ \hline
C (3) & Online Game & 500 & 8 & 50 & 0.72 & $10^{-3}$\\ \hline
D (80) & VR/AR & 1250 & 2 & 10 & - & $10^{-6}$\\ \hline
E (7) & Video Streaming & 1250 & 10 & 100 & - & $10^{-3}$\\ \hline
F (8) & FTP & 1250 & 6 & 300 & - & $10^{-2}$\\ \hline
\end{tabular}
\end{table}

\begin{table}
\renewcommand{\arraystretch}{1.3}
\caption{Scenarios in Various UE Traffic Type Ratios}
\label{table:traffic_scenarios}
\centering
\footnotesize
\begin{tabular}{|c|l|}
\hline
\textbf{Scenario} & \textbf{A : B : C : D : E : F} \\
\hline
Scenario 1 & 1 : 1 : 1 : 1 : 1 : 1 (Balanced)\\ \hline
Scenario 2 & 1 : 1 : 1 : 2 : 1 : 1 (Double VR/AR traffic)\\ \hline
Scenario 3 & 1 : 1 : 1 : 1 : 1 : 2 (Double FTP traffic)\\ \hline
Scenario 4 & 3 : 3 : 3 : 1 : 1 : 1 (More UEs with GBR)\\ \hline
Scenario 5 & 1 : 1 : 2 : 2 : 1 : 1 (More UEs with short latency)\\ \hline
Scenario 6 & 1 : 1 : 1 : 2 : 1 : 2 (More UEs with high data rate)\\ \hline
\end{tabular}
\end{table}

\begin{table}
\renewcommand{\arraystretch}{1.3}
\caption{Communication System Parameters}
\label{table:system_parameters}
\centering
\footnotesize
\begin{tabular}{|c|c|}
\hline
\textbf{Parameter} & \textbf{Value} \\
\hline
Operating Frequency & 3.5GHz\\ \hline
Number of Antennas per BS ($M$) & 128 \\ \hline
Total Number of UEs ($K$) & 100 to 600 \\ \hline
Average Number of Co-existing UEs ($K_r$) & 20 \\ \hline
Bandwidth ($B$) & 20 MHz \\ \hline
Transmit Power ($\rho$) & 24 dBm\\ \hline
Transmission Time Interval (TTI, $T_I$) & 1 ms\\ \hline
Channel Model & COST2100 \\ \hline
Antennas Array Type & Cylindrical Array \\ \hline
UEs Distribution & Poisson Point Process \\ \hline
Cell Range & radius of 100m \\ \hline
Termination Time ($T$) & 60000 TTI \\ \hline
\end{tabular}
\end{table}

\begin{table}
\renewcommand{\arraystretch}{1.3}
\caption{DDPG Parameters}
\label{table:ML_parameters}
\centering
\footnotesize
\begin{tabular}{|c|c|}
\hline
\textbf{Parameter} & \textbf{Value} \\
\hline
Replay Buffer Size & 600000 TTI \\ \hline
Mini Batch Size & 60000 TTI \\ \hline
Actor Learning Rate & $2e^{-3}$ \\ \hline
Critic Learning Rate & $1e^{-3}$ \\ \hline
Reward Discount, $\gamma$ & 0.9 \\ \hline
Soft Replacement $\tau$ & $1e^{-2}$ \\ \hline
Dropout Rate & 0.5 \\ \hline
Explore Rate & 0.1 \\ \hline
Hidden Layer & (512, 512, 512) \\ \hline
Optimizer & Adam \\ \hline
\end{tabular}
\end{table}

Several fundamental method combinations and algorithms in the literature are compared with the proposed one to evaluate dynamic method selection effectiveness with DDPG. The benchmark actions are the most frequently selected ones from the learning results and are kept static during simulation runs. For example, CQI-MinG75-AS applies channel quality first, minimum guarantee with 75\% resources reserved, and antenna selection precoder. We expect the learning scheme to choose a suitable resource allocation combination under various traffic and channel status of the environment. The static schemes always apply the same algorithms.

Related joint resource allocation works, including ORFA\cite{Singh2016}, UBLAA\cite{Kuo2018}, and LWDF-PF\cite{song2005cross,Girici2010}, are also implemented for comparison. Due to variations in system setups, we extract algorithms from the works to apply in our environment with the proposed concepts maintained. The selected algorithms have to cover user prioritization and are compatible with the antenna allocation process. Also, if necessary, we supplement comparing algorithms with best-performing precoding methods to form fair evaluation.
The ORFA specified all three function components: proportional fair UE ranking, water filling resource allocation, and linear MMSE precoding.
The UBLAA defines the marginal utility in a massive MIMO video streaming system to prioritize UEs for antenna allocation. The AS precoder, which matches the greedy-based UBLAA algorithm, is applied.
The LWDF-PF is a landmark proportional fair algorithm that adopts weighted delay fairness for UE scheduling; it serves as a general baseline. We choose the ACE precoder, which performs the best among $\mathbf{C}_3$, for LWDF-PF.

\subsection{Dynamic vs. Static Algorithm Combinations}

We compare the proposed learning-based method against static combinations in this section to demonstrate the advantages of dynamic algorithm combinations. Performance metrics in total system utilities~\eqref{eq:problem}, which is the proposed main objective, and throughputs are illustrated.

\begin{figure}
\centering
\subfloat[Normalized System Utility.]{\includegraphics[width=8.5cm]{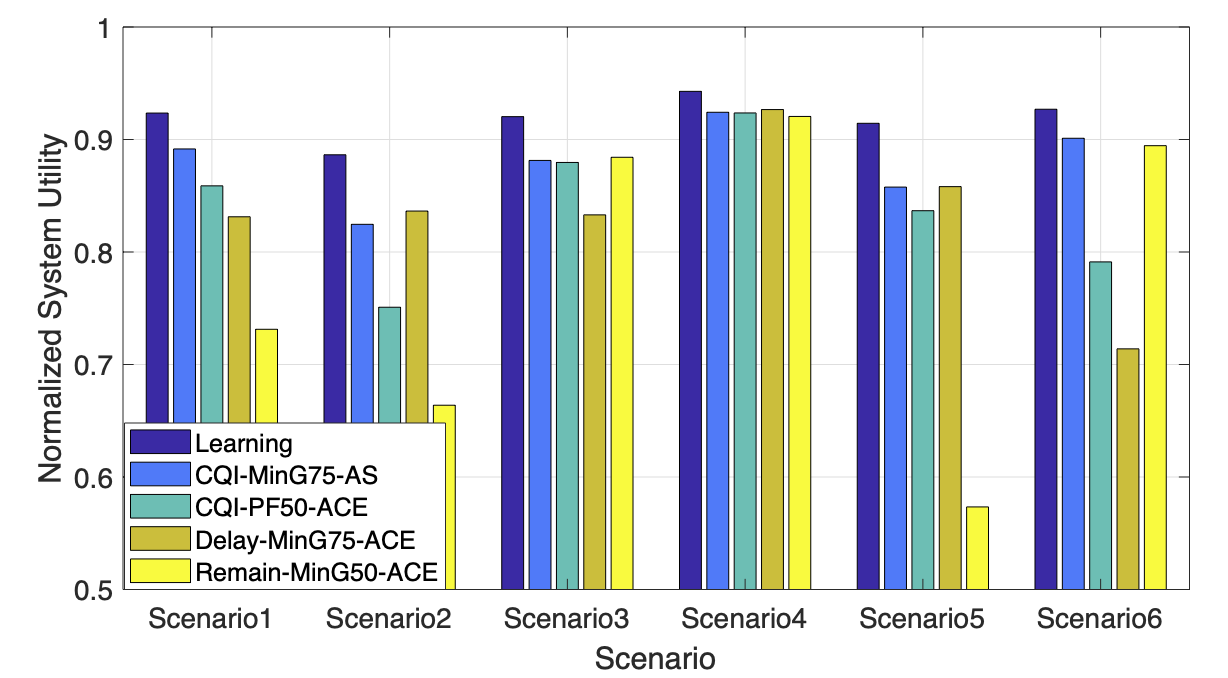}
\label{fig:satis_single}}
\hfil
\subfloat[Throughput]{\includegraphics[width=8.5cm]{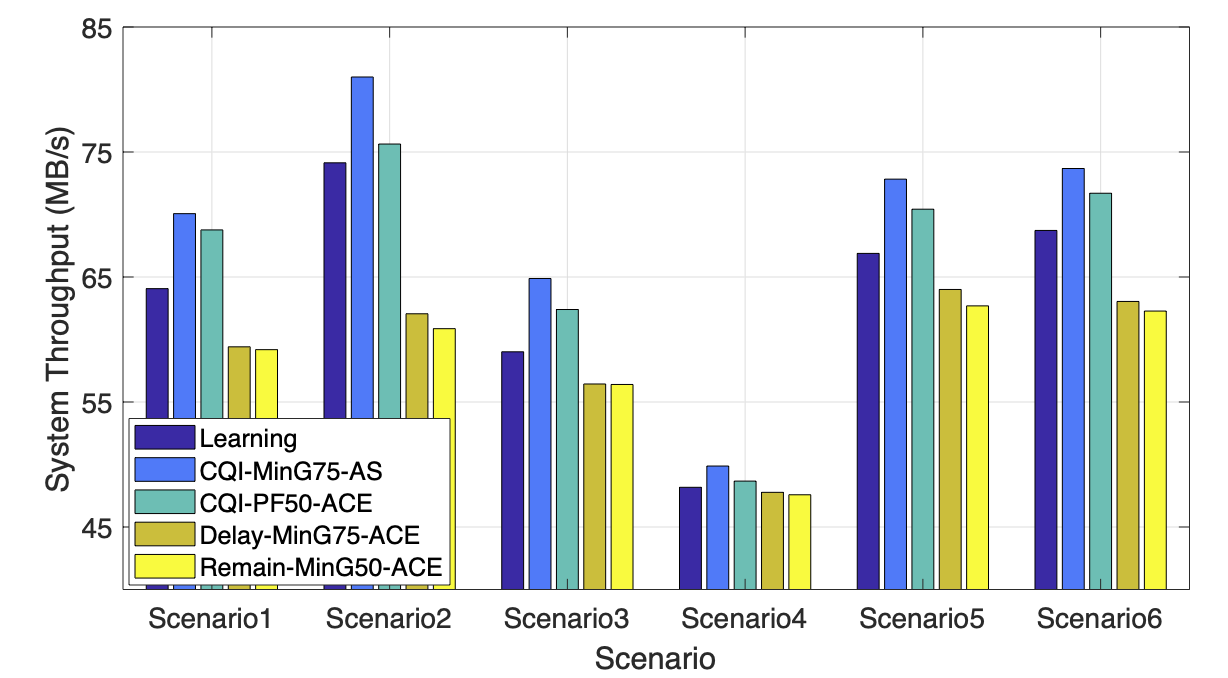}
\label{fig:throughput_single}}
\caption{Compare with static methods.}
\label{fig:learning_vs_static}
\end{figure}

Figure~\ref{fig:satis_single} illustrates the normalized system utility defined as the percentage of satisfied UEs through termination time $T$. We observe that the proposed learning-based approach gains 2.2\% to 7.2\% more system utility than the best static scheme across all scenarios due to its adaptive nature. The most significant advantage appears in Scenario 2 with doubled VR/AR traffic showing the learning method is more capable of achieving high bandwidth and low latency simultaneously. The performances of static schemes are inconsistent across application scenarios.
For example, the delay emphasizing scheme, Delay-MinG75-ACE, ranks second in Scenario 2 and 5, where more latency demanding VR/AR or gaming traffic exists. The scheme Remain-MinG50-ACE is comparable with the best ones in data rate demanding Scenario 3, 4, and 6, but achieves significantly less in others because UEs with more remaining data are ranked higher. Furthermore, CQI-MinG75-AS is a more versatile static combination because CQI provides high system throughput while MinG75 forces even distribution of most antenna resources. The greedy nature of AS precoder is also fitted well with CQI and MinG75.

From the system throughputs presented in Figure~\ref{fig:throughput_single}, we observe that greater throughput not necessarily reflects greater utility. Schemes that apply the CQI method for $c_1$, CQI-MinG75-AS and CQI-PF50-ACE, result in the highest throughputs because UEs are ranked according to channel quality. The proposed learning-based method ranked only behind CQI methods in throughput and outperforms them in system utility. When the overall traffic demand and throughput are lower in Scenario 4, all schemes achieve system utility greater than 0.9.

\begin{figure}
    \centering
    \includegraphics[width=8.5cm]{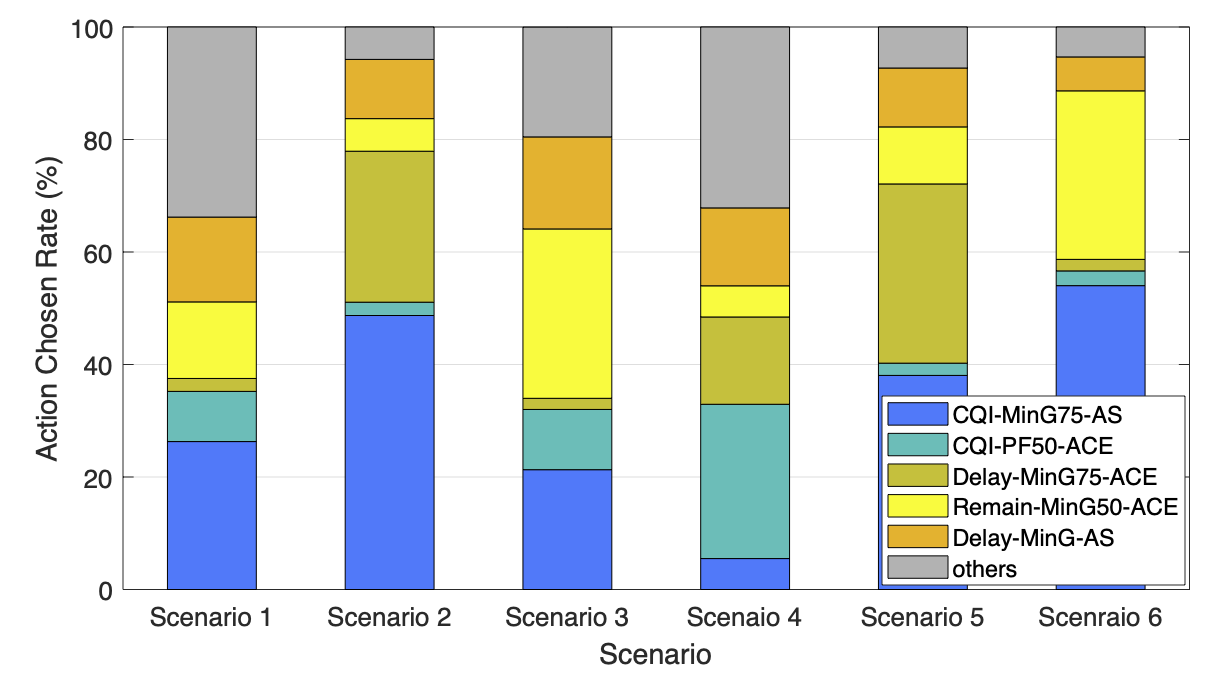}
    \caption{Action choose rate in DDPG agent}
    \label{fig:action_choose}
\end{figure}

Algorithm selection details in Figure~\ref{fig:action_choose} can further reveal the advantage of the learning method. The figure breaks down the proportion of top actions chosen by the DDPG agent in simulated scenarios. We observe that the learned strategies are adjusted accordingly to the scenarios. In Scenario 2 and 5, where the proposed method gains more than 7\% utility than others, CQI-Min75-AS and Delay-MinG75-ACE are most frequently selected with more than 70\% TTIs combined. It clearly shows that switching between throughput and delay priority timely is an effective strategy to fulfill challenging throughput and latency requirements simultaneously. In Scenario 3 and 6, CQI-Min75-AS and Remain-MinG50-ACE are applied the most for high data rate applications. When the overall demand is low in Scenario 4, CQI-PF50-ACE is applied more to emphasize proportional fairness. In balanced Scenario 1, most action combinations other than the top five are selected. Overall, the simulations demonstrate the effectiveness of forming an advanced resource allocation solution by adaptively switching between fundamental methods. The proposed componentized action structure is the key to realize this concept under advanced DRL training.

\subsection{Comparison with Joint Resource Allocation Algorithms}

\begin{figure}
\centering
\subfloat[Normalized System Utility]{\includegraphics[width=8.5cm]{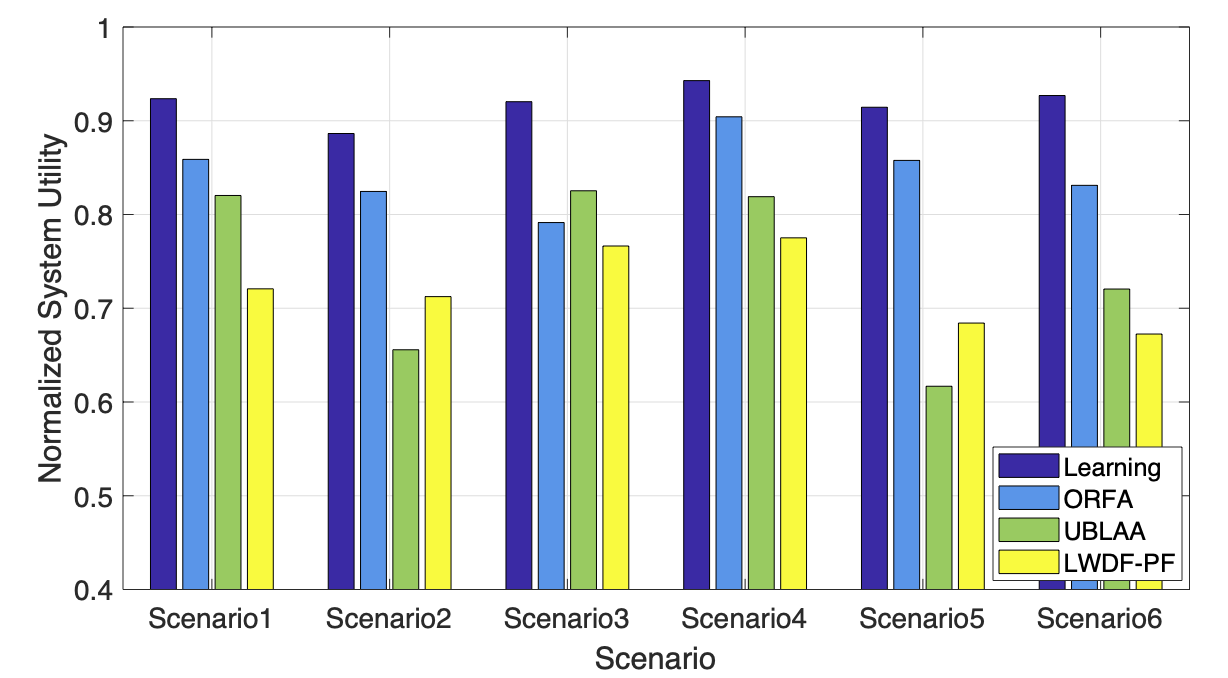}
\label{fig:satis_other}}
\hfil
\subfloat[Throughput]{\includegraphics[width=8.5cm]{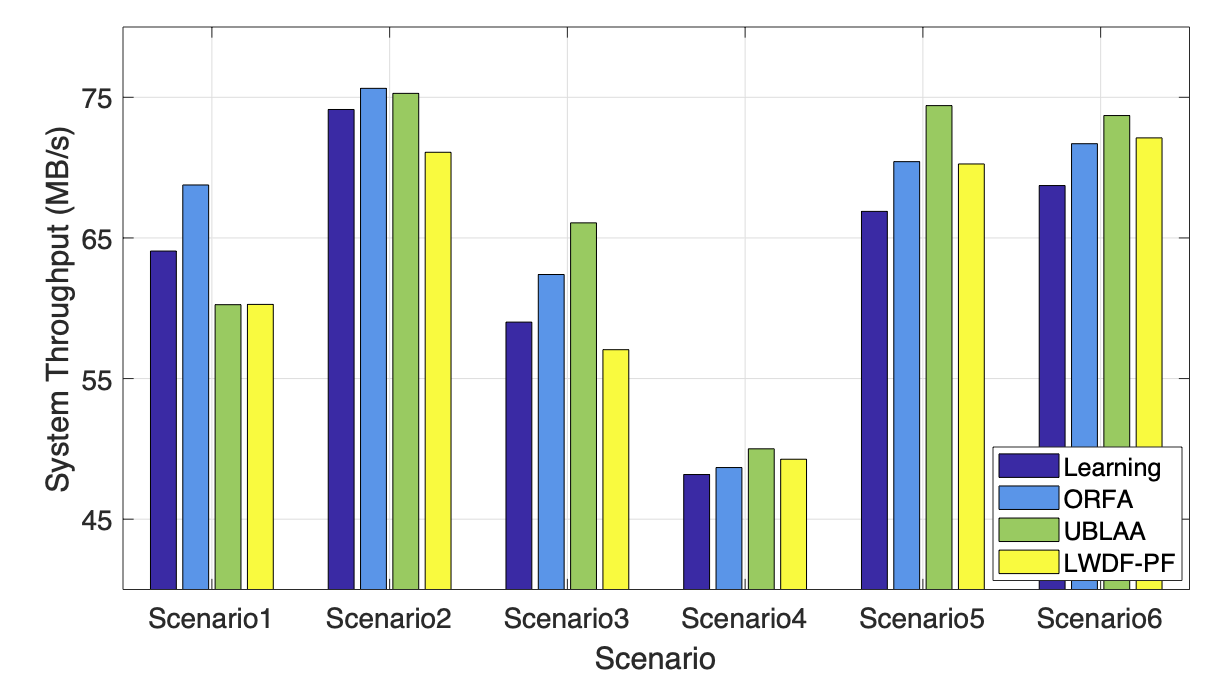}
\label{fig:throughput_other}}
\caption{Compare with other methods.}
\label{fig:learning_vs_others}
\end{figure}

Figure~\ref{fig:satis_other} and~\ref{fig:throughput_other} present the normalized system utility and system throughput comparing with joint resource allocation algorithms in the literature. Though not providing the highest throughputs, the proposed learning approach outperforms the best among ORFA, UBLLA, and LWDF-PF algorithms in normalized system utility. The largest utility gap is observed at 12.5\% in Scenario, with heavy traffic on high data rate and low latency types. In comparison, the smallest gap presents in the less loaded Scenario 4 at 4.4\%.
ORFA consistently achieves greater than 0.8 in utility due to the optimality of the water-filling algorithm. However, its' general-purpose proportional fair scheduling suffers from degraded performance under diverse application requirements.
UBLAA fulfills data rate requirements and results in high throughput in all scenarios.  Since latency is not effectively presented via marginal utility, system utility performance is not satisfactory in Scenario 2, 5, and 6, with latency demanding VA/AR applications.
LWDF-PF performs worse than the learning and ORFA methods and does not suffer from significant drops by adopting the relatively straightforward greedy weighted delay and proportional fairness allocation.

\begin{figure*}
    \centering
    \subfloat[CDF of short-term average utility for Scenario 1]{\includegraphics[width=5.7cm]{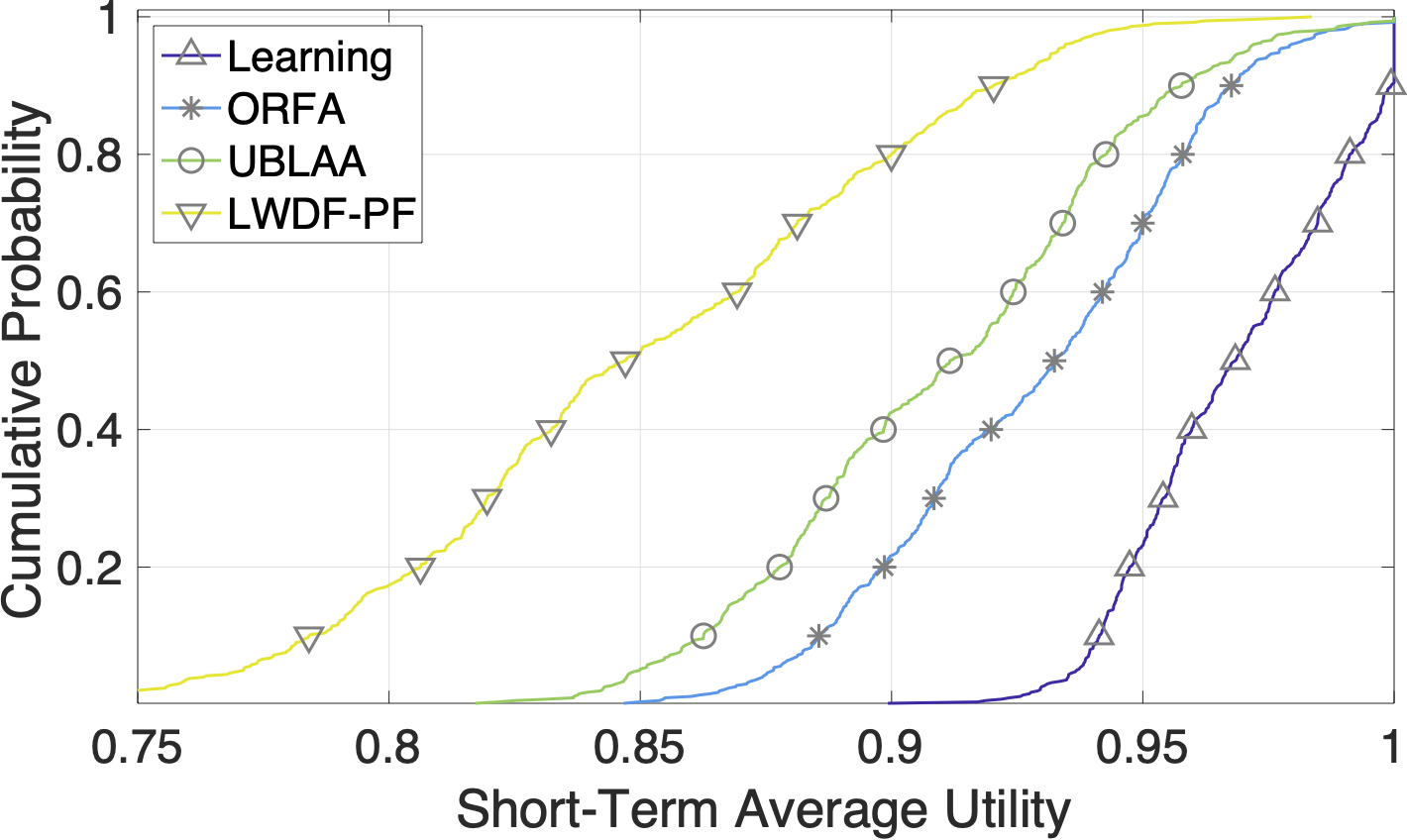}
    \label{fig:cdf_s1}}
    \hfil
    \subfloat[System utility vs. number of antennas for balanced case]{\includegraphics[width=5.7cm]{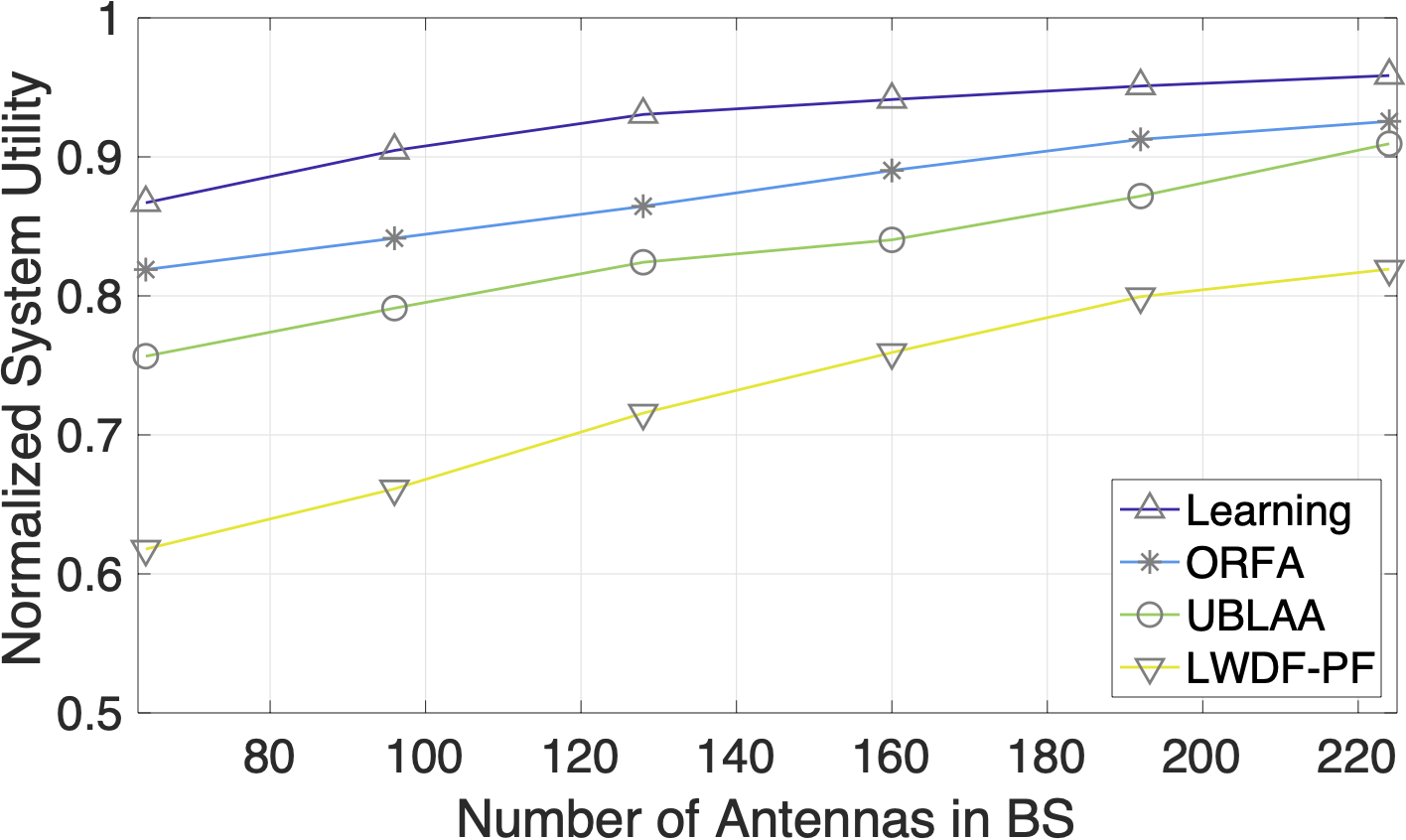}
    \label{fig:antennas_s1}}
    \hfil
    \subfloat[System utility vs. number of UEs for balanced case]{\includegraphics[width=5.7cm]{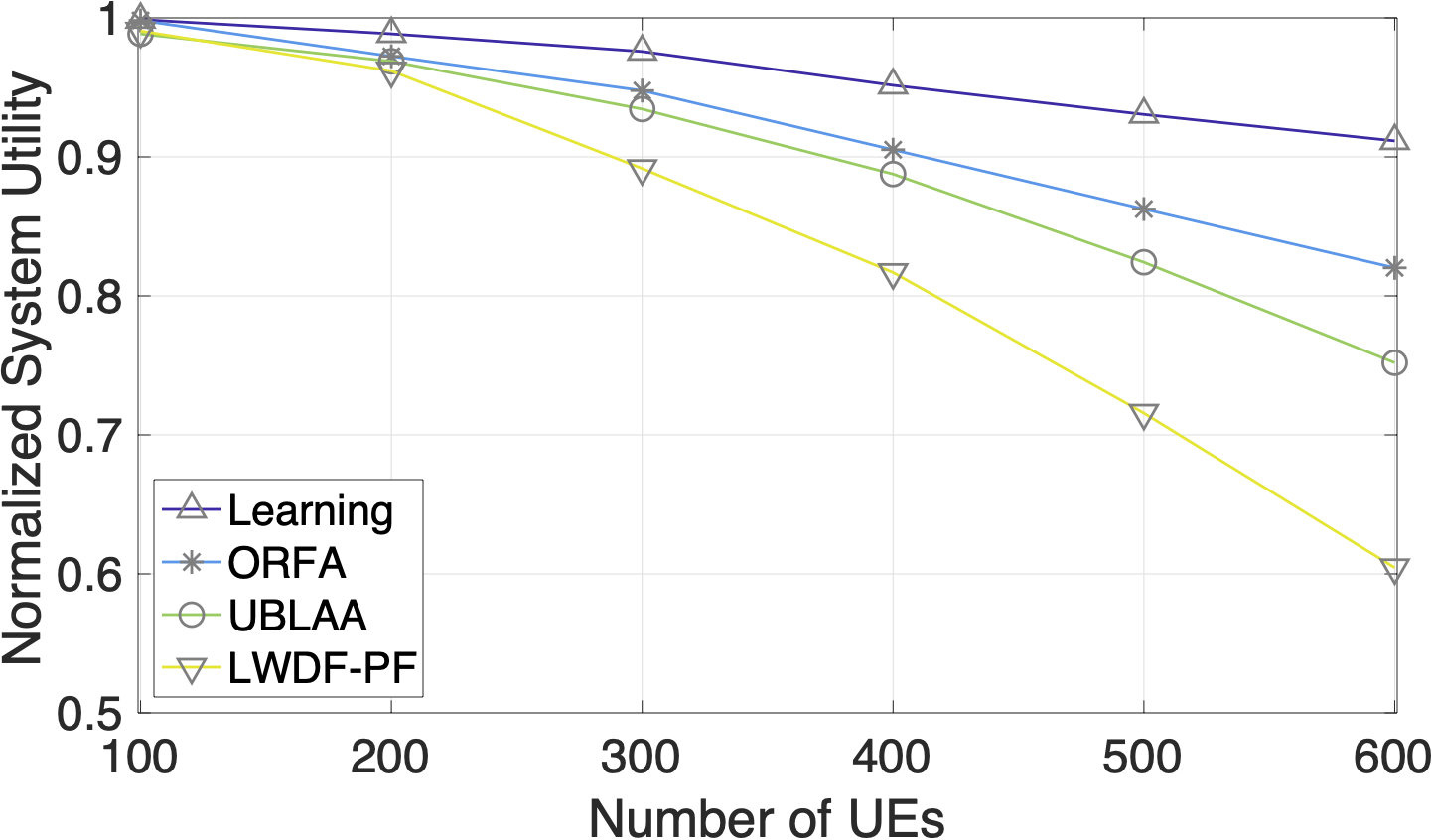}
    \label{fig:UEs_s1}}

    \subfloat[CDF of utility for AR/VR emphasized case]{\includegraphics[width=5.7cm]{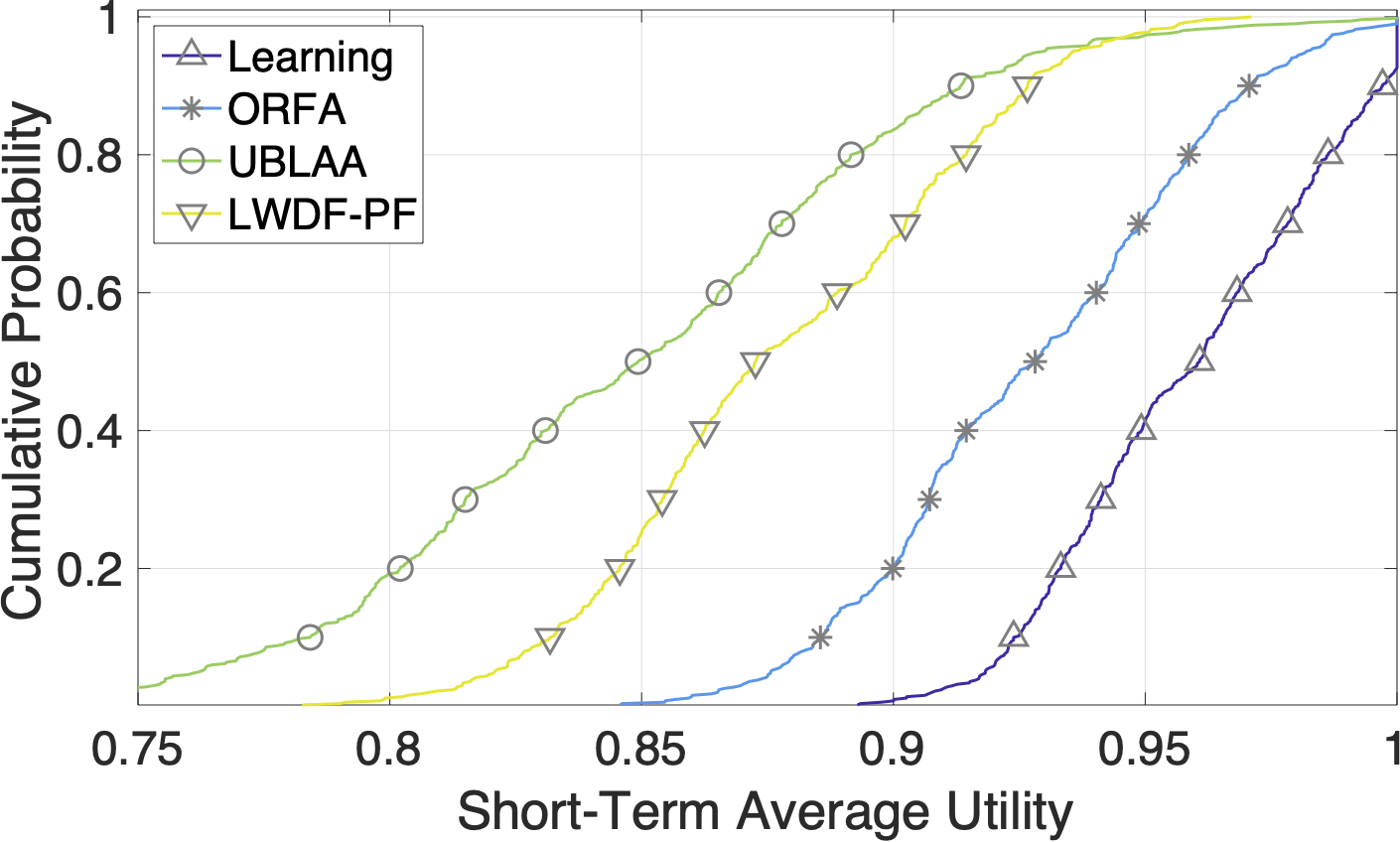}
    \label{fig:cdf_s2}}
    \hfil
    \subfloat[System utility vs. number of antennas for AR/VR emphasized case]{\includegraphics[width=5.7cm]{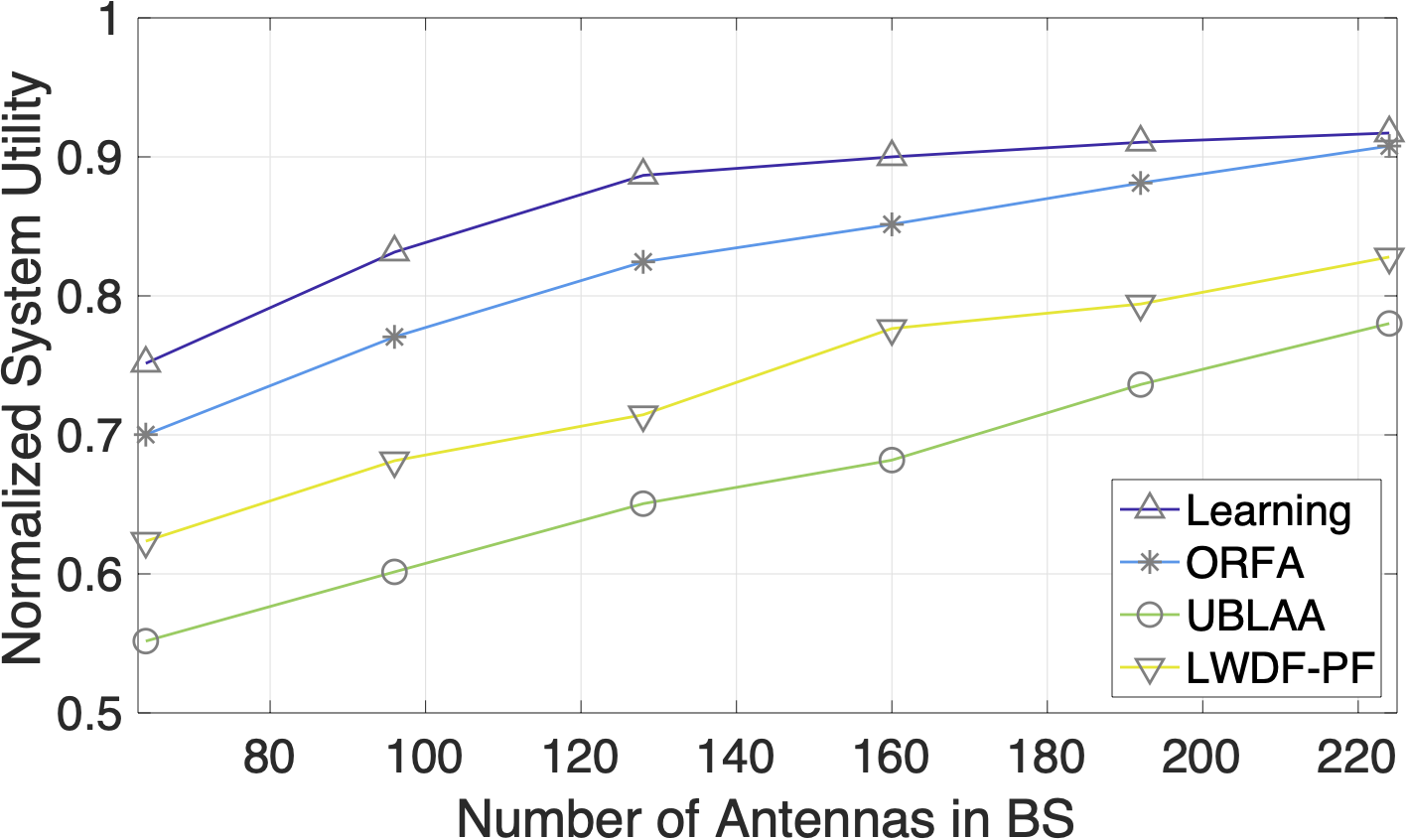}
    \label{fig:antennas_s2}
    }
    \hfil
    \subfloat[System utility vs. number of UEs for AR/VR emphasized case]{\includegraphics[width=5.7cm]{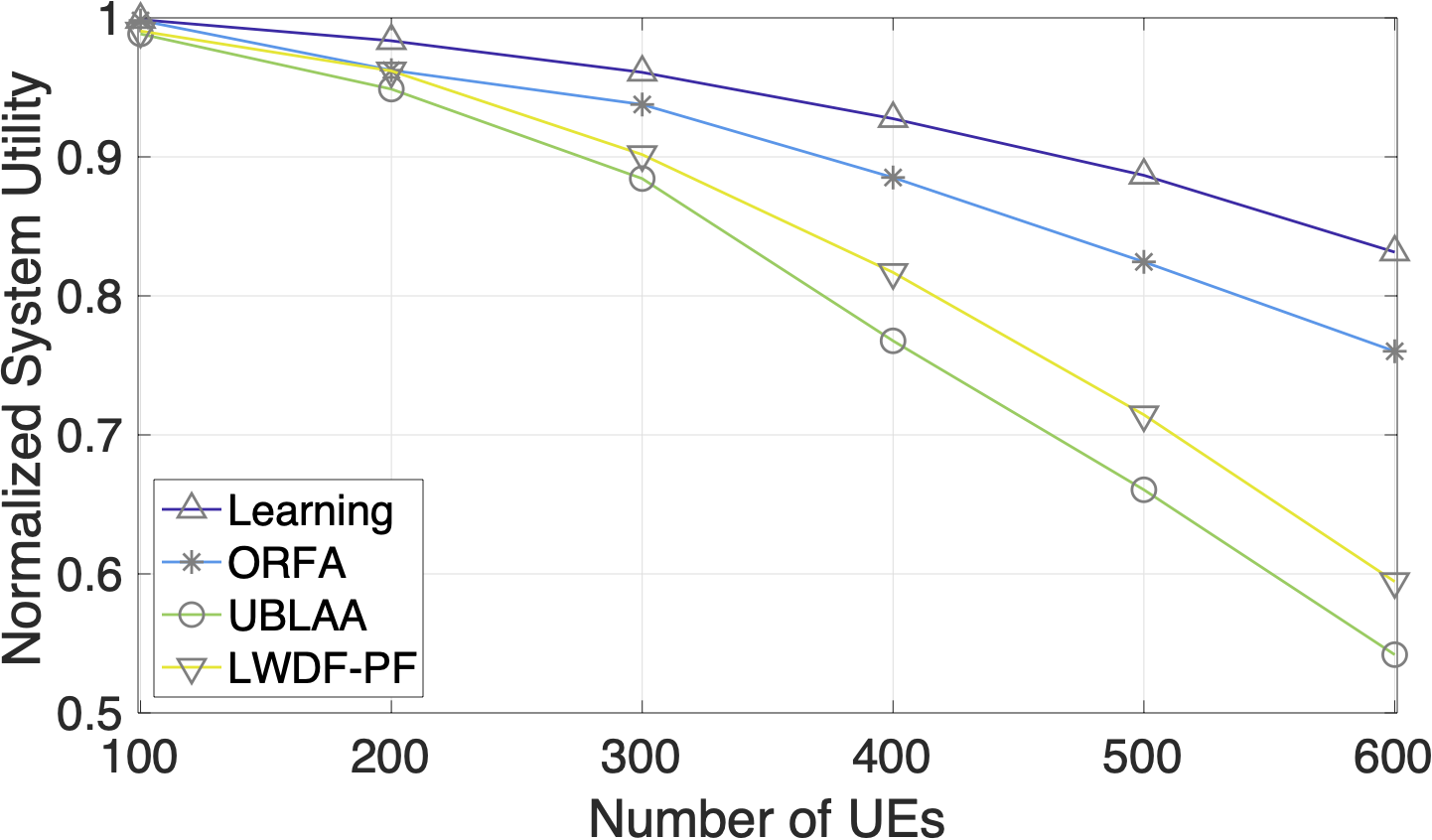}
    \label{fig:UEs_s2}
    }
    \label{fig:s1_vs_s2}
    \caption{Simulation results for specific cases.}
    \label{fig:specific_scenarios}
\end{figure*}

Figure~\ref{fig:specific_scenarios} shows detailed results for two representative scenarios. Scenario 1 with a balanced traffic mixture and Scenario 2 emphasizing VR/AR applications are selected. In the proposed problem \eqref{eq:problem}, the overall utility of a UE, $U_k$, can be concluded after the last requested data is processed at the session ending TTI of $k$. The average utility of UEs ending in the same 100-TTI windows is evaluated as \emph{short-term average utility} to analyze the system condition over time. 
Figures~\ref{fig:cdf_s1} and~\ref{fig:cdf_s2} illustrate the cumulative distribution function (CDF) of short-term average utilities with 128 antennas and 500 UEs. We can see that the learning method spread mainly in 0.9 and above. ORAF has samples lower from 0.85. UBLLA and LWDF-PF perform differently in Scenario 1 and 2. UBLLA keeps all samples greater than 0.83 in the balanced condition, while some samples fall below 0.75 when there is more latency-sensitive traffic in the system like Scenario 2.
Figure~\ref{fig:antennas_s1} and~\ref{fig:antennas_s2} present the system utility trend using 64 to 224 BS antennas with 500 total (42 average coexist) UEs. In general, systems gain more utility with more antennas. When the resources are limited at 64 antennas, the learning method gains 6.2\% to 40\% more utility than others in the balanced cases and 7.1\% to 22\% more in VR/AR emphasized cases.
Figure~\ref{fig:UEs_s1} and~\ref{fig:UEs_s2} show the system utility trend with 100 to 600 total (8.3 to 50 average coexist) UEs at 128 antennas. The advantage of learning-based algorithm selection grows with the saturation level resulting from a greater number of UEs. Also, overall utilities drop faster in VR/AR emphasized Scenario 2.

To summarize, the comparing joint methods fulfill the user scheduling and resource allocation problem objective~\eqref{eq:scheduling_problem} in general, where the decision is made to maximize the instant utility $U_{k,t}$. In contrast, the proposed MDP-based method maximizes the long-term utility~\eqref{eq:utility} and thus joint objective~\eqref{eq:problem}, because maximizing the long-term return~\eqref{eq:return} is the nature of MDP. The cross-layer integration of scheduling and precoding also shows effectiveness.

\section{Conclusion}\label{sec:conclusion}

A DRL-based radio resource allocation approach for joint scheduling and precoding in a massive MIMO system is investigated in this work. We suggest an architecture decomposing the cross-layer adaptation decision as a combination of algorithms and learning a dynamic algorithm selection policy in challenging 5G traffic scenarios. Comprehensive simulations are carried out to justify the effectiveness of the proposed method. Overall, the componentized structure can be the core of an extensible smart agent to deal with complex decision-making problems in future mobile networks.









\bibliographystyle{IEEEtran}
\bibliography{ref}

\vfill

\end{document}